\documentclass[printer]{aa}

\usepackage{txfonts}
\usepackage{graphicx}
\usepackage{natbib}
\usepackage{mathrsfs}
\usepackage{url}
\bibpunct{(}{)}{;}{a}{}{,}

\begin{document}

\title{
Cores, filaments, and bundles: hierarchical core formation in the
L1495/B213 Taurus region
\thanks{Based on observations carried out with the FCRAO 14m and IRAM 30m 
telescopes. 
IRAM is supported by INSU/CNRS (France), MPG (Germany) and IGN (Spain). 
Also based on data acquired with the Atacama Pathfinder Experiment (APEX).
APEX is a collaboration between the Max-Planck-Institut fur Radioastronomie,
the European Southern Observatory, and the Onsala Space Observatory
(ESO projects 080.C-3054 and 083.C-0453).
}}

\titlerunning{Hierarchical core formation in L1495/B213}

\author{A. Hacar 
\inst{1,2}
\and
M. Tafalla 
\inst{1}
\and
J. Kauffmann
\inst{3}
\and
A. Kov\'acs
\inst{4}
}

\institute{Observatorio Astron\'omico Nacional (IGN),
Alfonso XII 3, E-28014 Madrid,
Spain
\and
Institute for Astrophysics, University of Vienna, T\"urkenschanzstrasse
17, 1180 Vienna, Austria. \\
\email{alvaro.hacar@univie.ac.at}
\and
Jet Propulsion Laboratory, California Institute of Technology, 4800 Oak Grove 
Drive, Pasadena, CA 91109, USA
\and
University of Minnesota, 116 Church St SE, Minneapolis, MN 55414, USA
}
\offprints{A. Hacar}
\date{Received -- / Accepted -- }

\abstract
%context
{Core condensation is a critical step in the star-formation process,
but is still poorly characterized observationally.
}
%aims
{We have studied the 10~pc-long L1495/B213 complex in Taurus to  
investigate how dense cores have condensed out of the lower-density
cloud material.
}
%methods
{We have observed L1495/B213 in C$^{18}$O(1--0), N$_2$H$^+$(1--0),
and SO(J$_{\mathrm{N}}$=3$_2$--2$_1$) with the 14-m FCRAO telescope,
and complemented the data with dust continuum observations 
using APEX (870~$\mu$m) and IRAM 30m (1200~$\mu$m). 
}
%results
{
From the N$_2$H$^+$ emission, we identify 
19 dense cores, some starless and some protostellar. They are 
not distributed uniformly, but tend to cluster with 
relative separations on the order of 0.25~pc.
From the C$^{18}$O emission, we identify multiple velocity
components in the gas. We have characterized them
by fitting gaussians to the spectra, and by studying
the distribution of the fits in position-position-velocity
space. In this space, the C$^{18}$O components appear as 
velocity-coherent
structures, and we have identified them automatically
using a dedicated algorithm 
(FIVe: Friends In Velocity). 
Using this algorithm, we have identified 
35 filamentary components with typical lengths
of 0.5~pc, sonic internal velocity dispersions,
and mass-per-unit-length 
close to the stability threshold of isothermal
cylinders at 10 K. 
Core formation seems to have occurred
inside the filamentary components via fragmentation, with a small 
number of fertile components with larger mass-per-unit-length being
responsible for most cores in the cloud.
At large scales, the filamentary components appear grouped into families,
which we refer to as bundles.
}
% conclusions
{
Core formation in L1495/B213 has proceeded by hierarchical
fragmentation. 
The cloud fragmented first into several pc-scale regions.
Each of these regions later fragmented 
into velocity-coherent filaments of about 0.5~pc in length.
Finally, a small number of these
filaments fragmented quasi-statically and 
produced the individual dense cores we see today.
}


\maketitle

%
%________________________________________________________________

\section{Introduction}

Dense cores are the sites of individual or binary stellar birth
\citep{ben89, dif07}, and their condensation from the ambient cloud
represents a critical step in the process of star formation.
Core formation, however, appears as a bottleneck in
the road from clouds to stars.
At any given time, only a small fraction of the material in a cloud 
is in the form of dense cores ($< 10$~\%, i.e., \citealt{eno07}), and this
apparent difficulty making dense cores is
likely connected to the low efficiency
of the star-formation process \citep{eva08}.
Core formation, in addition, seems to play a key role
in determining the mass of the final star, since the
distribution of masses among  starless cores in a cloud
mimics the initial distribution of stellar masses
\citep{mot98,alv07}.
Core formation, therefore, represents
a  crucial gas transition
by which a cloud selects a small fraction of its material to form 
the next generation of stars and leaves the
rest as a sterile remnant
to be dispersed into the more diffuse
interstellar medium.

Despite its critical role in star-formation, the process
of core formation is still poorly understood.
A number of condensation mechanisms
have been proposed over the years, 
ranging from quasi-static contraction mediated by ambipolar-diffusion
\citep{shu87,mou99}
to gravo-turbulent fragmentation driven by supersonic motions
\citep{pad01,kle05,vaz05}.
Observations, however, do not favor clearly any single model,
partly due to the intrinsic difficulty 
in measuring time scales of core evolution 
and magnetic field intensities, which are the
main parameters that distinguish the different models
(\citealt{war07} for a review).

New approaches to study core formation are needed, and
a promising strategy is to 
analyze in detail the velocity field of the gas,
comparing the internal motions
of the dense cores with those of the surrounding 
low-density gas from which they have condensed.
Following this approach, \citet{hac11} found that the 
gas surrounding the cores of the L1517 cloud 
is subsonic and quiescent, like
the gas inside the dense cores, in contradiction
with the predictions from models of gravo-turbulent fragmentation.
More interestingly, the lower-density gas in L1517
forms a network of filaments whose velocity fields are 
continuous and subsonic
over scales of about 0.5~pc, which is
significantly larger than the typical core size ($\approx 0.1$~pc).
These so-called velocity-coherent filaments
seem therefore to constitute the parent structures
from which the cores form. The
observed continuity between the large-scale velocity field of the filaments
and the internal velocity gradients of the cores indicates
that the transition between the two regimes
involves little kinematical changes, and in particular,
an absence of shock compression.

The analysis of L1517 suggests 
that core formation in this Taurus cloud 
has occurred in two steps, with the velocity-coherent
filaments forming first 
and the cores fragmenting later from the already-quiescent filament gas.
In this scenario, turbulent dissipation precedes core formation, and
gravitational fragmentation of the filaments is the final step in the 
core-formation sequence.

Independent work using dust continuum data from the Herschel
Space Observatory has revealed that filamentary structures in clouds
are ubiquitous, and that filament-based core formation is likely 
to be a widespread process
in both low and high mass star-forming regions \citep{and10, mol10,arz11}.
(See also \citealt{sch79,joh99,har02,mye09} for previous work
emphasizing the importance of filaments in core formation.)
Understanding core formation inside filaments has therefore become an 
urgent task, and kinematic information is a needed element to
complete the picture of filament physics.
For this reason, we have selected the most prominent filamentary
region of the Taurus cloud, which consists of the B211, B213, B216, B217, and 
B218 dark patches \citep{bar27} and ends in the L1495 cloud 
\citep{lyn62}, and we have subjected it to observations similar
to those used to study L1517.

Our cloud of study,  which we will refer to as L1495/B213 for brevity,
appears in optical images as a several-degree-long 
filament first noticed by \citet{bar07}, who incidentally used
this and other Taurus filaments to argue that at least some dark objects 
result from the obscuration by a physical {\em substratum,}
and not from the mere absence of background stars as previously thought.
Due to its striking appearance, the
L1495/B213 complex has been investigated and characterized 
by a number of authors over the years.
Initial extinction maps were presented by \citet{gai84}
and \citet{cer85} using optical data, and more recently by \citet{sch10}
using deep NIR observations.
The gas component of the cloud has been studied systematically
in $^{13}$CO and C$^{18}$O by \citet{duv86},
who noticed a complex velocity structure that was
interpreted as resulting from  colliding filaments.
Additional large-scale maps of the region in CO isotopologues
have been presented by \cite{hey87}, \citet{miz95}, \citet{oni96}, and 
\citet{gol08}. Very recently, \citet{pal13} have presented high dynamic
range Herschel images of the dust continuum emission.

Embedded in the relatively low-density CO-emitting gas of L1495/B213,
lies a population of dense cores, which has been explored by 
\citet{ben89} and \citet{oni02} using NH$_3$ and H$^{13}$CO$^+$ observations,
respectively, with additional and more limited N$_2$H$^+$ observations
presented by \citet{tat04}. 
Some of these dense cores are starless, while
other are associated with young stellar objects (YSOs) of different
ages. This population of YSOs has been the subject of
a number of dedicated studies, most recently by
\citet{luh09} and \citet{reb10}, and by the dedicated 
outflow search from  \citet{dav10}.

The above observations, 
and additional work reviewed by 
\citet{ken08}, show that the L1495/B213 complex
has been and still is an active site of
star formation in Taurus. In fact, all
stages of the star-formation sequence can be found 
in the cloud,
from starless cores that may have recently condensed, to
Class 0 and Class I protostars with active outflows and 
signs of accretion, to classical and weak T Tauri stars
in their path to the main sequence. This rich
population of objects
makes L1495/B213 an ideal laboratory to study core
and star-formation, with the additional advantage that the region has 
a well-defined filamentary geometry that simplifies its mapping.

\section{Observations}\label{obs}

The main dataset used in this paper consists of observations
carried out with the FCRAO 14m radio telescope
during several sessions between March 2002 and November 2005. 
In each session, the telescope was equipped with the
32-pixel SEQUOIA focal-plane array receiver and the 
DCC autocorrelator, which allowed us to observe
two different spectral lines simultaneously, either in the
85-100 GHz or the 100-115 GHz bands. 
As the two main lines of the project, N$_2$H$^+$(1--0) and
C$^{18}$O(1--0), could not be observed simultaneously, 
the L1495/B213 region was mapped twice.
In one pass,  N$_2$H$^+$(J=1--0) and SO(J$_{\mathrm{N}}$=3$_2$--2$_1$)
were observed simultaneously,
and in the other pass, 
C$^{18}$O(J=1--0) and C$^{17}$O(J=1--0)
were observed.
To achieve high velocity resolution, 
the DCC autocorrelator was configured to provide
1024 spectral channels with a spacing of 25~kHz, or
approximately 0.07~km~s$^{-1}$.
Rest frequencies are assumed to have the same values as
in \citet{hac11}.

All observations were done in on-the-fly mode, covering the cloud
with a mosaic of $10'\times 10'$ maps
referred to a common center at 
$\alpha(J2000)= 4^h17^m47\fs1,$ 
$\delta(J2000)= +27^\circ37'18''$.
Position switching was used to subtract the sky and receiver
contributions, and the
reference position was located ($+2300''$, $-4500''$) with
respect to the map center. 
This position has a C$^{18}$O(J=1--0)
peak intensity $T_{mb}<0.03$~K averaged over 
the SEQUOIA footprint, as estimated from frequency switched observations.
This intensity is negligible compared to the typical 
on-source values, which are larger than 1~K. 

During the observations, calibration was done every 
10 minutes, and pointing corrections were determined every
3 hours from SiO maser observations, finding typical errors
within $5''$ rms. According to FCRAO-provided information
(\url{http://www.astro.umass.edu/~fcrao/observer/status14m.html}),
the beam efficiency of the telescope is approximately 0.5, and the
FWHM beam size depends linearly on frequency, having a value
of $56''$ at the lowest (N$_2$H$^+$)
frequency and of $47''$ at the highest (C$^{17}$O)
frequency.

Data reduction consisted in the creation of Nyquist-sampled maps
with the {\tt otftool} program
(\url{http://www.astro.umass.edu/~fcrao/library/manuals/otfmanual.html}),
and resulted in data sets of
more than 35,000 spectra for each of our four program lines.
These spectra were later converted into the GILDAS/CLASS format 
(\url{http://www.iram.fr/IRAMFR/GILDAS}) for 
second-degree baseline subtraction and a
spatial convolution with a gaussian
to eliminate residual noise.
The final resolution of all the FCRAO data,
taking into account the off-line convolution, is $60''$,
and the grid spacing is $30''$. With this sampling, 
the size of each molecular
dataset is approximately 23,000 spectra.

Additional line observations of L1495/B213 were made using
the IRAM 30m telescope. These complementary data
consisted of simultaneous N$_2$H$^+$(1--0) and C$^{18}$O(2--1) spectra 
observed toward a selected group of positions to clarify the
velocity structure of the gas. The observations were done in
2010 April and 2011 December using the EMIR heterodyne receiver
in frequency switching mode. The backend was
the VESPA autocorrelator configured to provide
a velocity resolution of 0.06~km~s$^{-1}$. Sky calibration was
carried out every 15 minutes, and pointing corrections every 
2 hours. Conversion to the mean brightness temperature
scale was done using the facility-provided telescope
efficiencies. The angular resolution of the observations 
was approximately $25''$ for N$_2$H$^+$(1--0) and $12''$
for C$^{18}$O(2--1).

Dust continuum observations of selected parts of L1495/B213
were carried out at 1200~$\mu$m with the MAMBO-2 array 
on the IRAM 30m telescope during several
runs between 2003 December and 2009 November.
The observations were done in on-the-fly
mode with a scanning speed of $8''$~s$^{-1}$,
a wobbler period of 0.5 s, and wobbler throws
between $50''$ and $70''$. Atmospheric calibration was
carried out using data from sky dips every
1.5 hours,
and the absolute calibration was achieved 
using observations of CRL618. 
The data were reduced with the MOPSIC software
using a method to recover extended emission consisting
of 10 iterations with a model source and subtraction
of correlated noise.

Additional dust continuum observations of L1495/B213 were carried out
at 870~$\mu$m with the LABOCA array on the APEX telescope
in 2007 November and 2009 July.
The observation consisted of two on-the-fly maps 
(2007 observations) and a mosaic of seven 
raster-spiral (quarter) maps \citep{sir09}
(2009 observations). Calibration was carried
out using sky dips and observations of Mars and CRL618
about every 2 hours, and the reduction used the BOA software
together with the method to recover extended emission presented
by \citet{bel11}. An artistic rendition of these observations 
can be found in the ESO photo release eso1209
(\url{http://www.eso.org/public/news/eso1209/}).

\section{Large-scale cloud properties from the integrated maps:
evidence for sequential fragmentation}\label{sect-large}

\begin{figure*}
\begin{center}
\resizebox{14cm}{!}{\includegraphics{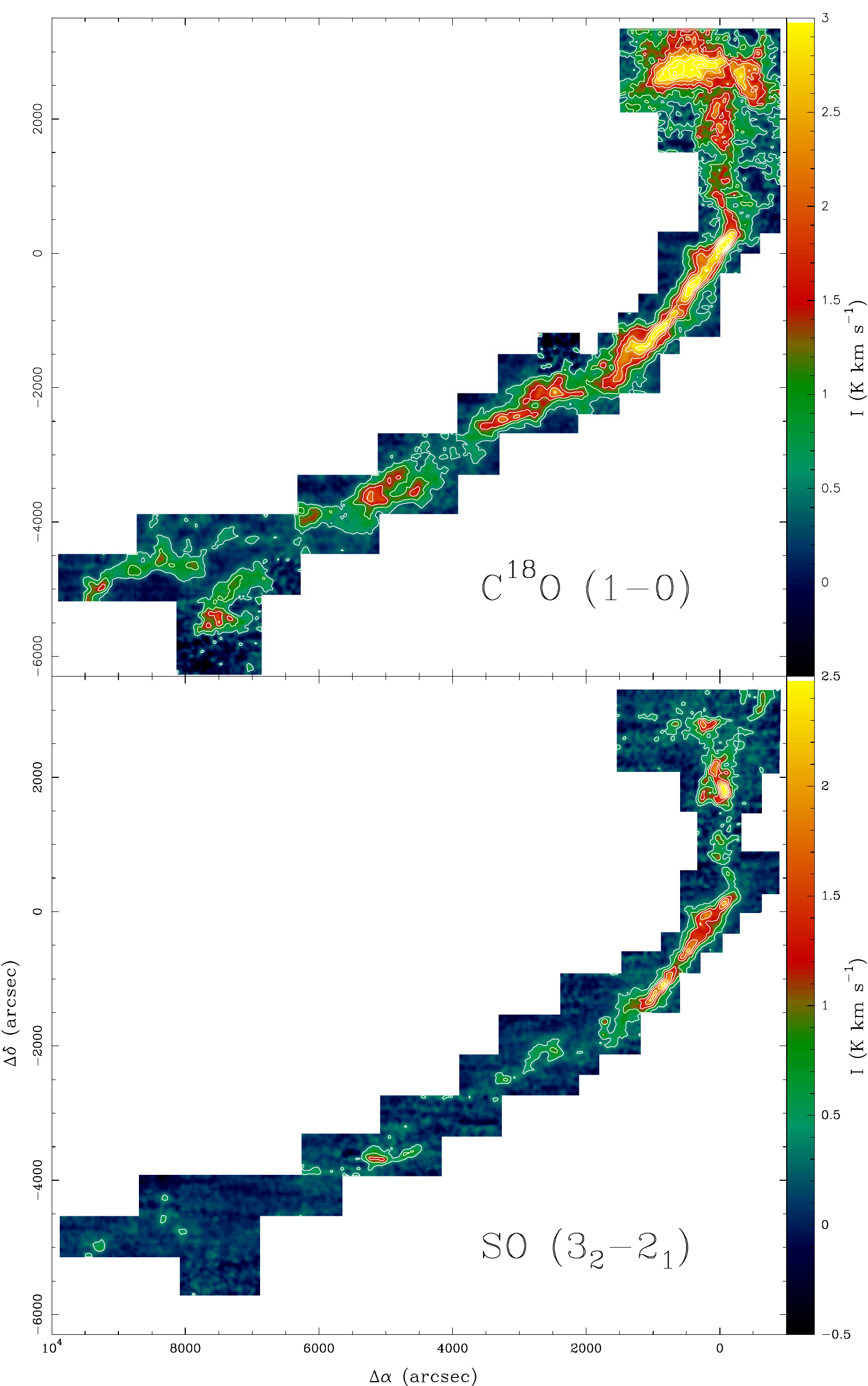}}
\caption{Integrated intensity maps of L1495/B213 in 
C$^{18}$O(1--0) (top) and SO(J$_{\mathrm{N}}$=3$_2$--2$_1$)
(bottom).
The offsets are referred to the FCRAO map center at 
$\alpha(J2000)= 4^h17^m47\fs1,$
$\delta(J2000)= +27^\circ37'18''$, and 
the maps have been convolved to a resolution of $75''$ to 
enhance sensitivity.
First contour and contour interval are
0.5~K~km~s$^{-1}$ for C$^{18}$O(1--0)
and 0.4~K~km~s$^{-1}$ for SO(3$_2$--2$_1$). 
\label{fcrao-large-1}}
\end{center}
\end{figure*}

\begin{figure*}
\begin{center}
\resizebox{14cm}{!}{\includegraphics{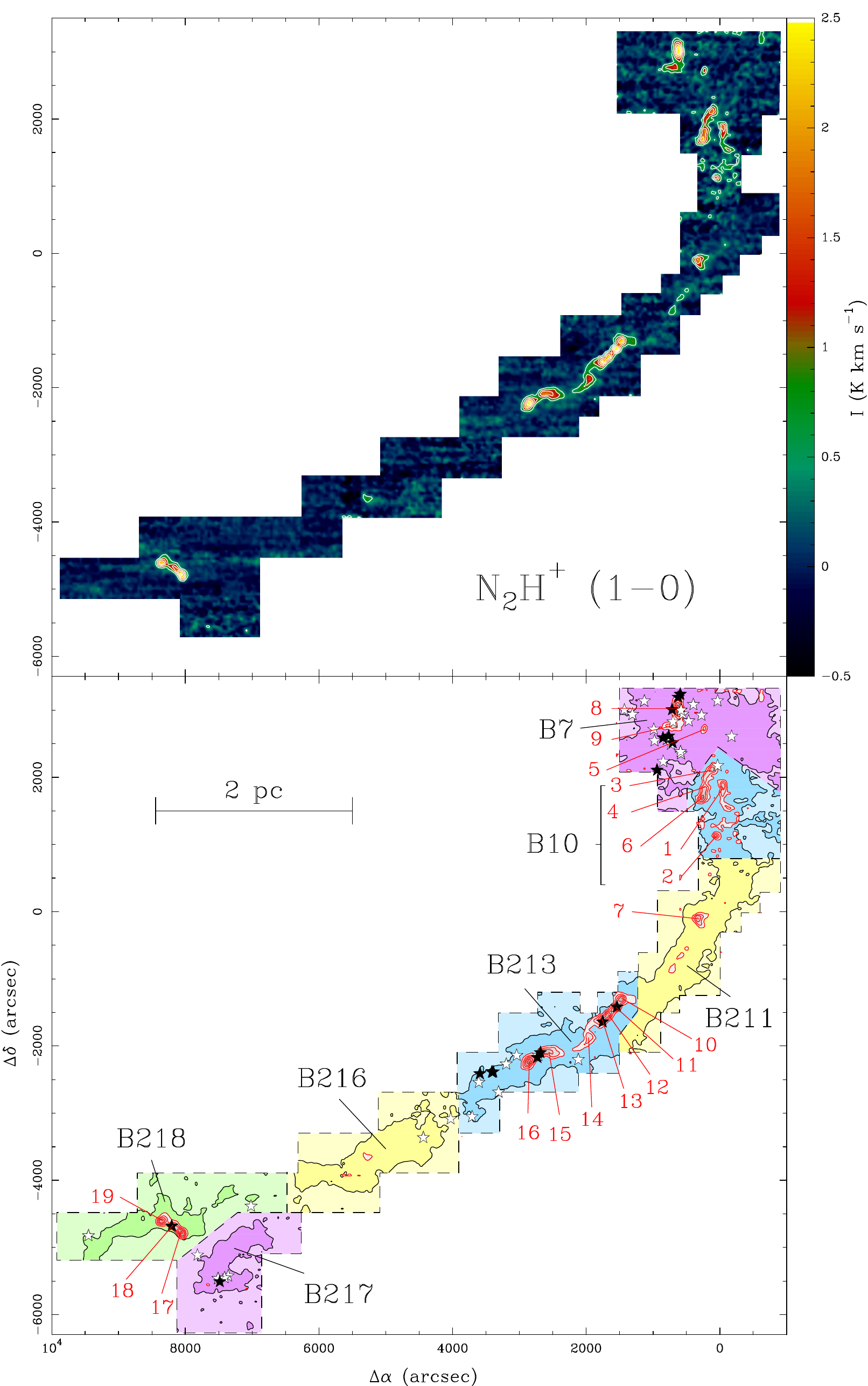}}
\caption{{\em Top:} Integrated intensity map of L1495/B213 in
N$_2$H$^+$(1--0) adding all hyperfine components. 
The white contour corresponds to 0.5~K~km~s$^{-1}$.
Central position and offsets are as in Fig.~\ref{fcrao-large-1}.
First contour and contour interval are 0.5~K~km~s$^{-1}$.
{\em Bottom:} Schematic view of the  L1495/B213 complex 
indicating the regions defined in this paper, which match
approximately those of \citet{bar27}.
The black solid line represents the lowest  C$^{18}$O(1--0) contour in  
Fig.~\ref{fcrao-large-1} and delineates the boundary of the
cloud. The red lines represent the 
N$_2$H$^+$(1--0) emission, which traces the dense cores,
and the red labels identify the cores
described in Sect.~\ref{sect-cores} and summarized in 
Table~\ref{tbl_cores}. 
The star symbols correspond to 
stellar objects from the survey of \citet{reb10}. Solid symbols
represent the youngest objects (Class~I and Flat) and open symbols
represent evolved objects (Class~II and Class~III).
\label{fcrao-large-2}}
\end{center}
\end{figure*}

Figs.~\ref{fcrao-large-1} and \ref{fcrao-large-2}
present integrated-intensity maps of
L1495/B213 in C$^{18}$O, SO, and N$_2$H$^+$.
The different species 
provide complementary information on the 
gas conditions due to a
combination of excitation and abundance effects.
C$^{18}$O is the most sensitive tracer of the low-density material
in the cloud thanks to its lower critical density and relatively large 
abundance, although it disappears from the gas phase in dense, chemically
evolved regions due to freeze out onto dust grains 
\citep{cas99,ber02,taf02}.
SO, on the other hand, presents a higher critical density that makes it
more selective of dense gas. Its abundance, however, 
is especially sensitive to the evolutionary state of the material, and is higher
during the earliest phases of gas contraction \citep{taf06,hac11}.
Finally, N$_2$H$^+$ is a tracer of dense, chemically evolved gas  due
to its abundance enhancement when CO freezes out, and thus 
highlights the population of dense cores in the cloud
(see \citealt{ber07} for a review on core chemistry).

As can be seen in Fig.~\ref{fcrao-large-1}, the 
C$^{18}$O emission (top panel) traces a curved filamentary cloud over
approximately $4^\circ$, or about 10~pc for our assumed distance
of 140~pc \citep{eli78}. This emission has a bright condensation towards the
northern end that coincides with the location of the L1495 dark cloud
and with one of the main groups of pre-main sequence stars in
Taurus \citep{ken08}. From L1495,
the emission extends first south and then south east, and contains
several elongated regions of enhanced intensity. Finally, it 
forks into two branches of relatively diffuse emission towards
the east and the south-east. 

Overall, the
distribution of C$^{18}$O emission matches 
the distribution of extinction 
derived by \citet{sch10} from NIR data with an angular resolution
similar to that of our FCRAO observations. 
It also matches approximately the LABOCA and MAMBO dust continuum
maps presented in the following section. There are however significant
departures between the C$^{18}$O emission and both the dust extinction and
emission maps. These departures coincide with the location of the dense 
cores that dominate the N$_2$H$^+$ emission, and they can be seen by comparing
the maps of C$^{18}$O and N$_2$H$^+$.
Between $\Delta\alpha$  = $1000''$  and $3000''$, for example, there is a
chain of several cores that is bright both in
N$_2$H$^+$ and the dust continuum 
(next section), but that presents only weak and diffuse  C$^{18}$O  emission
in the map of Fig.~\ref{fcrao-large-1}.
This anticorrelation between the C$^{18}$O and N$_2$H$^+$ emissions
is typical of regions of core formation, and indicates that CO
has depleted from the dense gas 
\citep{cas99,ber02,taf02}.
While CO depletion is common in the inner part of dense cores,
the region with the core chain in B213
is unusual in presenting depressed CO emission 
over a scale more than 0.5~pc in length.

While the large-scale geometry of L1495/B213 suggests
that the cloud is a single entity with a common physical origin,
the internal structure of the gas presents strong evidence for
fragmentation.
This can be seen in the maps of Fig.~\ref{fcrao-large-1}, and
it was already noticed by Barnard from the study of the optical images,
which lead to his sub-division of the cloud into
distinct condensations \citep{bar27}.
To better refer to the different parts of the cloud
in our study, we have divided the 
mapped area into regions, and we have labeled these regions
following the 
original notation by Barnard as closely as possible.
Barnard's description, however, is only qualitative, and some parts
of the cloud are not assigned to any of his condensations, so
the correspondence between our regions and Barnard's
regions should be considered only approximate (see  also \citealt{sch10}
for a similar attempt to sub-divide L1495/B213).
The resulting cloud division is shown in the bottom panel of
Fig.~\ref{fcrao-large-2} using
color-coded boxes superposed to the lowest
C$^{18}$O(1--0) contour from in Fig.~\ref{fcrao-large-1}.

Barnard's division of the cloud 
was purely based on the appearance of the optical images,
but is correlated with a number of independent cloud properties.
One of these properties is the stellar population.
In Fig.~\ref{fcrao-large-2}, we use star symbols to represent
the stellar objects classified as ``previously identified''
or ``most likely'' (A+ rank) Taurus members 
by the Spitzer telescope survey of \citet{reb10}.
Identifying the full stellar population of
this Taurus region is still subject to debate, especially
in the low luminosity end of the distribution 
(e.g., \citealt{pal12}), and
for this reason we have required
conservatively that our candidates have
an A+ rank in the notation of \citet{reb10}.
To distinguish between the different stellar classes, we
have used in Fig.~\ref{fcrao-large-2} solid  
star symbols for the  youngest objects (Class I and Flat,
see \citealt{reb10}), and open star symbols
for the more evolved objects (Class II and III).

As it can be seen, the stellar population of the cloud is not
distributed uniformly over the different Barnard regions.
Most YSOs are located in regions B7, B213, and B217, which
together contain 51\% of the cloud mass but have 88\%
of the stars (i.e., the other 49\% of mass only contains 
12\% of the stars, see the Table~\ref{tbl_masses}).
Even among these three regions, there are significant differences in the type of 
associated stellar objects.
B7 and B217 contain twice as
many evolved objects as young ones, while
the B213 region contains the same amount of young
and evolved objects.
These differences suggest 
that star formation has not occurred
simultaneously in all the regions, but
that different parts of the cloud 
have been forming stars at different rates.

Another property that correlates with 
the sub-division of the cloud in regions 
is the chemical composition of the gas.
Comparing the intensities of C$^{18}$O, SO, and N$_2$H$^+$
in Figs.~\ref{fcrao-large-1} and \ref{fcrao-large-2}, 
it is clear that the emission 
ratio for the different species 
changes between regions, even when taking
into account the different excitation requirements of the
tracers. The most striking contrast between regions occurs in
the contiguous B211 and B213 regions. 
As mentioned before, the B213 region
presents relatively weak
C$^{18}$O and SO emission while  is associated with bright
N$_2$H$^+$. This is indicative of strong CO and SO depletion, and
suggests that the gas in B213 is chemically evolved and has
condensed into a number of dense cores.
These two properties are in good agreement 
with the finding that B213 has
the highest fraction of young protostars.

\begin{table*}
\caption[]{Properties of the sub-regions in L1495/B213.\label{tbl_masses}}
\centering
\begin{tabular}{lcccccc}
\hline
\noalign{\smallskip}
Region  &   Mass$^{(1)}$        &       Cores$^{(2)}$        &     YSOs$^{(3)}$ &
I(SO)/I(C$^{18}$O)$^{(4)}$ & Cores/YSOs & I+F/II+III$^{(5)}$\\
         & (M$_\odot$)       &   &   \\
\noalign{\smallskip}
\hline
\noalign{\smallskip}
B7      & 205 & 3 & 25 & 0.18 & 0.12 & 0.4 \\
B10     & 70 & 5 & 1 & 0.42 & 5.0 & 0.0 \\
B211   & 138   & 1 & 0 & 0.40 & inf & - \\
B213   &  114  & 7 & 14 & 0.22 & 0.5 & 1.0\\
B216       & 83  &0 & 2 & 0.28 & 0.0 & 0.0\\
B217       &  43  & 0 & 5 & 0.13 & 0.0 & 0.3 \\
B218       &  52   & 3 & 3 & 0.23 & 1.0 & 0.5 \\
\hline
\end{tabular}
\begin{list}{}{}
\item (1) From C$^{18}$O assuming an abundance of $1.7 \times 10^{-7}$
and $T_{\mathrm K}= 10$~K;
(2) as determined in Sect.~\ref{sect-cores}; (3) as determined by
\citet{reb10}; (4) ratio of mean SO(3$_2$--2$_1$)
and C$^{18}$O(1--0) intensities; (5) ratio of Class~I+Flat sources
over Class~II+Class~III sources according to \citet{reb10}.
\end{list}
\end{table*}

In contrast with B213, the nearby B211 region presents
very bright emission in both C$^{18}$O and SO, together 
with intense dust millimeter continuum (see next section).
Its N$_2$H$^+$ emission, however,
is almost undetected by our observations, except for a single core
(number 7 in the Fig.~\ref{fcrao-large-2}) that is not associated with
the main part of the filament (see Fig.~\ref{cores-b211} below). 
This combination of bright C$^{18}$O and SO together with 
weak N$_2$H$^+$ indicates that the gas in B211
has an unusually young chemical composition. 
Such chemical youth 
is again in good agreement with the lack
of  stellar population,
which has already been used to argue
that the B211 region is at a very early state of evolution
\citep{gol08,sch10}.

While B211 presents the strongest evidence  for
chemical youth, B10
also seems to be relatively un-evolved.
This can be seen in
Figs.~\ref{fcrao-large-1} and \ref{fcrao-large-2}, where B10 presents
relatively bright SO emission, especially when compared
to the C$^{18}$O-brighter B7 region to
its north. In contrast with B211, however, B10 is relatively 
weak in C$^{18}$O and has a number of N$_2$H$^+$-bright
dense cores, which suggests that some CO depletion has already taken
place. B10 must therefore be somewhat more evolved 
than B211. Interestingly, while B10 has already formed
a number of dense cores, none of them seems associated
to any Class I or Flat objects in the catalog of \citet{reb10}.

In summary, the evolutionary state of the different Barnard regions
is as follows.  B211 appears to be the least evolved region
due to its chemical composition, lack of protostars,
and presence of a single (starless) core.
B10 seems to follow in terms of evolution,
since it has a number of dense 
cores but only one is possibly associated 
with a YSO.
B216 appears also to be at a similarly young stage, due to
its lack of protostars and cores, but its more diffuse
nature makes its status less clear.
Next in evolution comes B213, that has started 
recently forming stars and is doing so very actively.
Follows B218, which has a number
of starless cores, including the very young L1521E
\citep{hir02,taf04b},
and has a very young stellar object.
Finally, B7 and B217 have a number of Class~I/Flat objects,
but whose stellar population is dominated by more
evolved objects.
Among these two regions, B7 has the largest number of
Class~III objects (7 out of 25 YSOs), so it is likely
to be more evolved than the rest.

While the differences in evolutionary stage between the 
parts of the L1495/B213 complex are clear, the assignment of  relative
ages is more uncertain.
Overall, we can estimate that some parts of the cloud 
have been forming stars for at least 1-2~Myr.
B7, for example, has 7 Class~II and Class~III objects with ages
in the range 1-6~Myr, according to the estimates by \citet{bert07}.
B213 is associated with FS Tau B, with 
an estimated age of 2.5~Myr, and
B217 contains FV Tau/c, which seems older than 
1~Myr \citep{kra09}.
At the other end of the scale, the B211 region,
despite being more massive than B213 or B217, seems to have not
yet formed a single star.
This means that the large-scale cloud, despite its striking
appearance as a single 10~pc filament, has not fragmented as
a single entity, but that different parsec-sized
regions have followed different star-formation histories 
over the last 1-2~Myr. 

Since the differences in stellar population are correlated
with differences in the chemical composition of the gas 
(Table~\ref{tbl_masses}),
and since this composition is very sensitive to the amount 
of time the gas has remained dense (e.g., \citealt{ber07}),
it seems unlikely that regions with little or no star formation,
like B211, have been ``waiting'' with their current physical
conditions for more than 1-2 Myr, while regions like B217 were forming stars.
To appear chemically young, 
regions like B211 must have condensed more recently from
less-dense cloud material than regions like B217.
This suggests that the complex fragmentation of the cloud
reflects a similarly complex pattern of assemblage of the large-scale
filament, by which ``evolved'' regions like B7 and B217
condensed and started to form stars first, while other
regions like B211 seem to have become dense only very recently.
As we will see below, this pattern of differential
evolution has left an imprint in the large-scale
pattern of gas velocities.

\section{Dense cores in L1495/B213}
\label{sect-cores}

\begin{figure}
\begin{center}
\resizebox{\hsize}{!}{\includegraphics{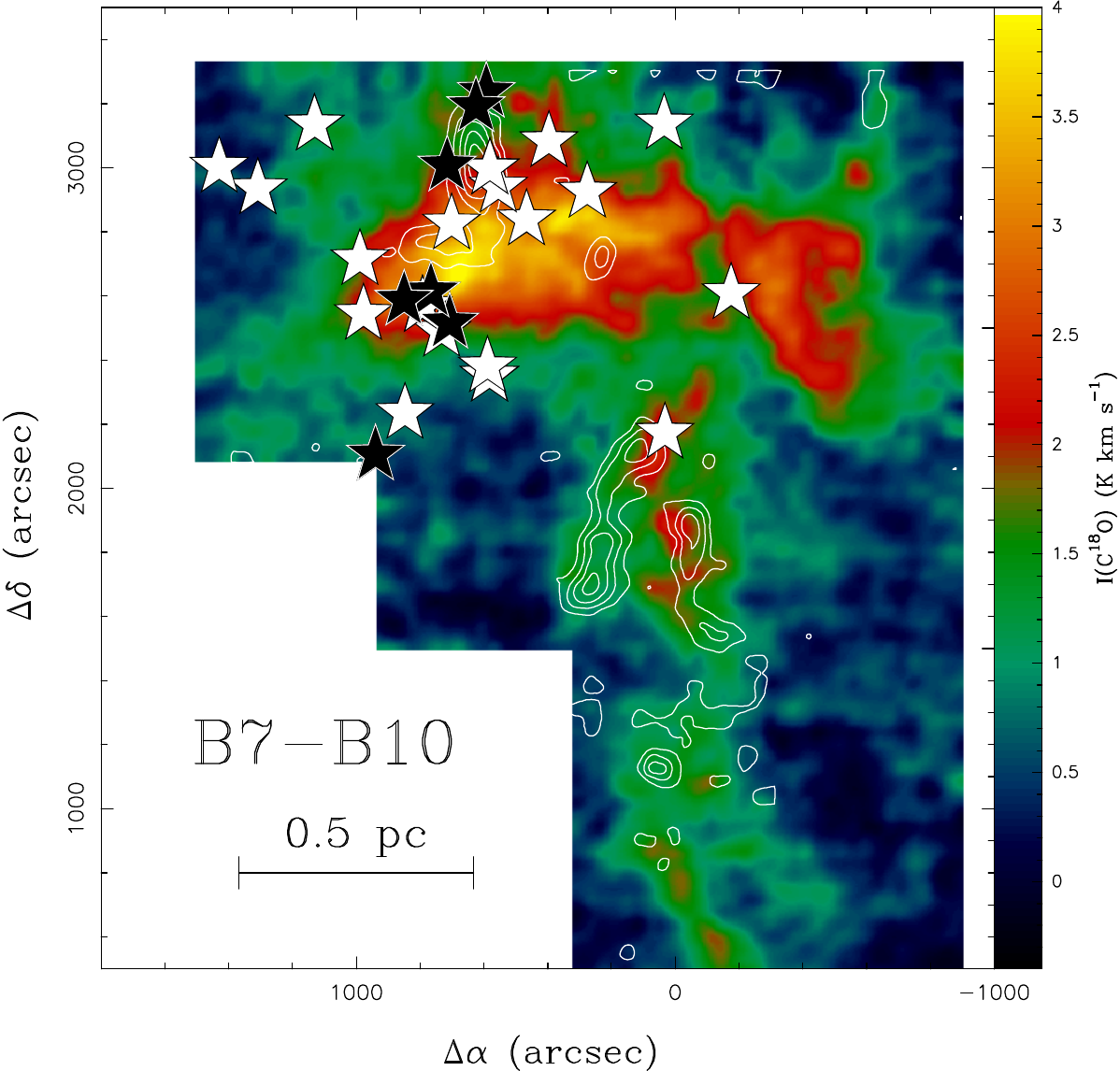}}
\caption{Expanded view of the B7-B10 region.
C$^{18}$O(1--0) emission in color and N$_2$H$^+$(1--0)
emission in white contours 
(sum of all hyperfine components).
First contour and contour interval are 0.5~K~km~s$^{-1}$.
Star symbols as in Fig.~\ref{fcrao-large-2} and offsets
referred to the FCRAO map center (see Sect.~\ref{obs}).
\label{cores-b10}}
\end{center}
\end{figure}

\begin{figure}
\begin{center}
\resizebox{\hsize}{!}{\includegraphics{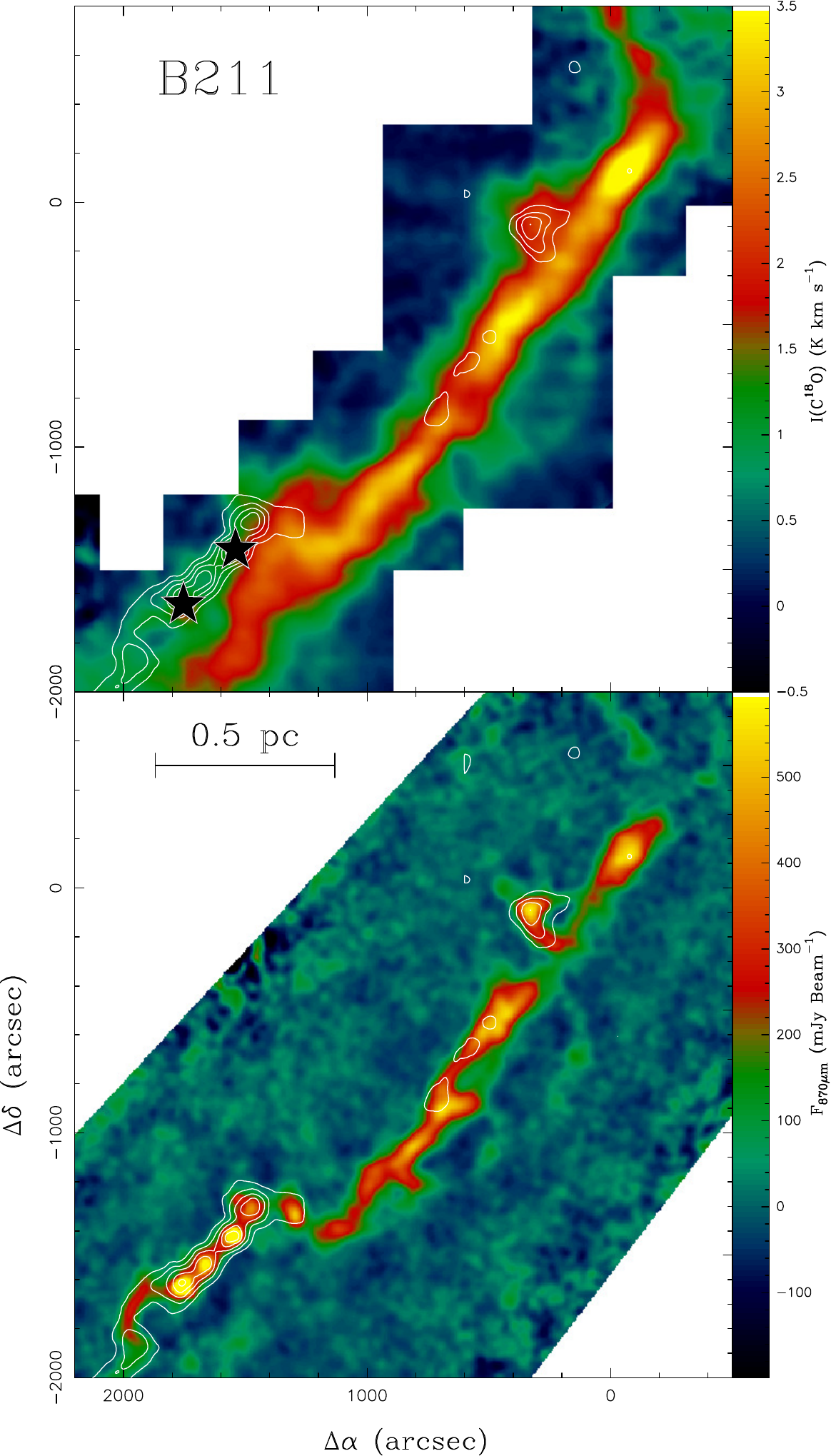}}
\caption{Expanded view of the B211 region and the NW part of B213 (to
better connect with the following figure).
{\em Top:} C$^{18}$O(1--0) and N$_2$H$^+$(1--0) maps as in Fig.~\ref{cores-b10}.
{\em Bottom:} 850~$\mu$m continuum map from LABOCA observations
and N$_2$H$^+$(1--0) emission (contours).
\label{cores-b211}}
\end{center}
\end{figure}

\begin{figure}
\begin{center}
\resizebox{\hsize}{!}{\includegraphics{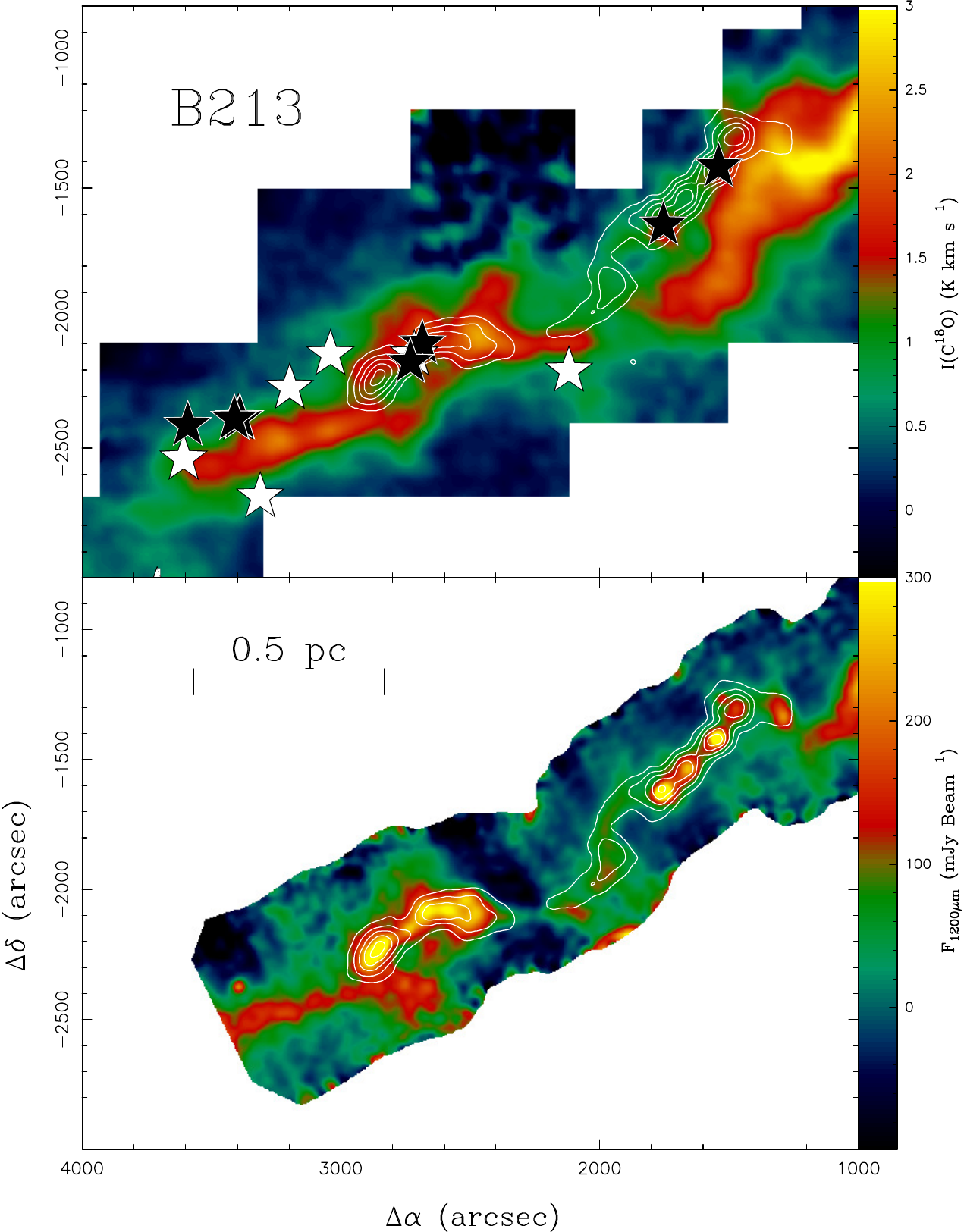}}
\caption{Expanded view of the B213 region.
{\em Top:} C$^{18}$O(1--0) and N$_2$H$^+$(1--0) maps as in Fig.~\ref{cores-b10}.
{\em Bottom:} 1200~$\mu$m continuum map from MAMBO observations
and N$_2$H$^+$(1--0) emission (contours).
\label{cores-b213}}
\end{center}
\end{figure}

\begin{figure}
\begin{center}
\resizebox{\hsize}{!}{\includegraphics{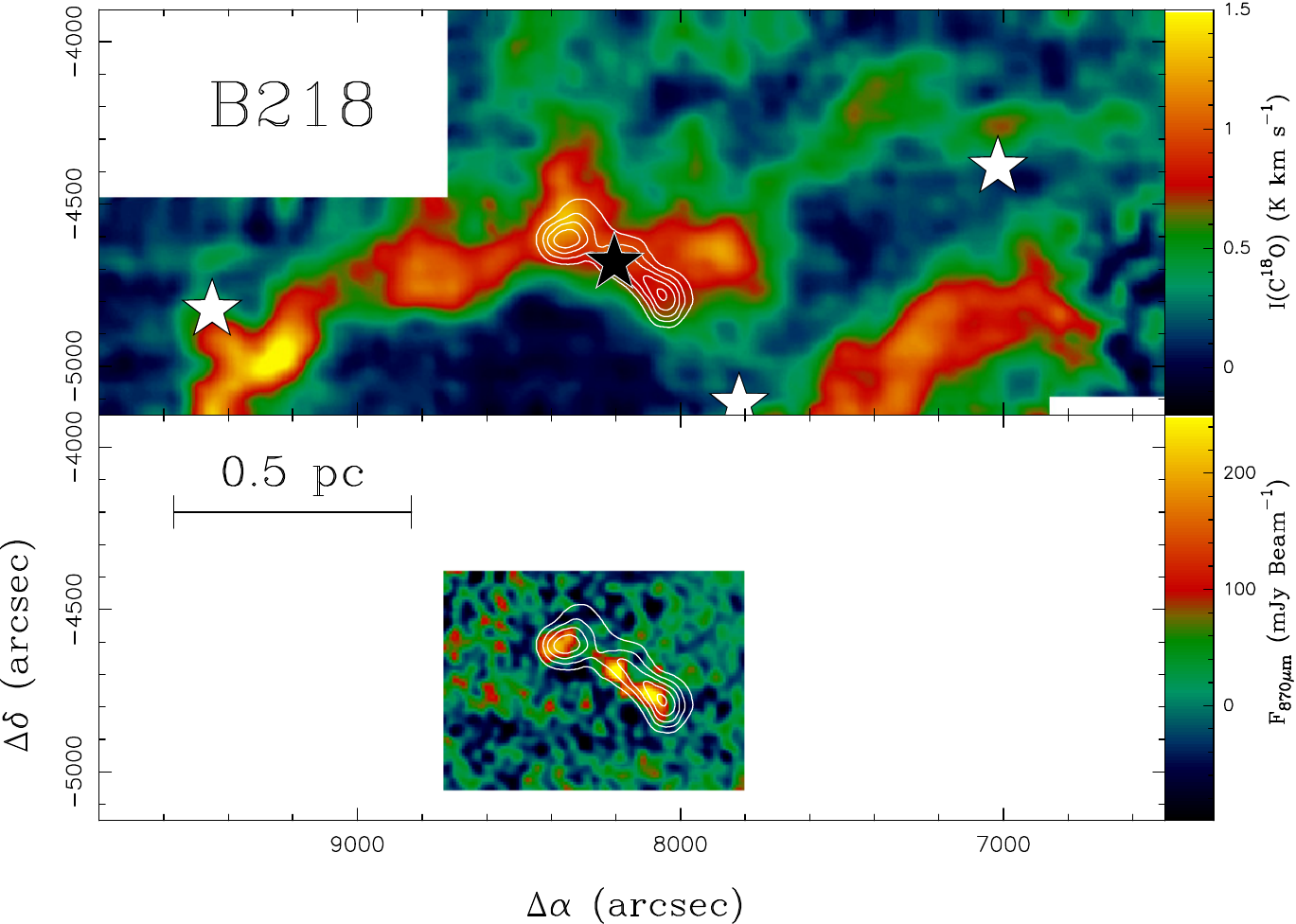}}
\caption{Expanded view of the B218 region. 
{\em Top:} C$^{18}$O(1--0) and N$_2$H$^+$(1--0) maps as in Fig.~\ref{cores-b10}.
{\em Bottom:} 850~$\mu$m continuum map from LABOCA observations
and N$_2$H$^+$(1--0) emission (contours).
\label{cores-b218}}
\end{center}
\end{figure}

The maps in Figs.~\ref{fcrao-large-1} and \ref{fcrao-large-2} 
are too large to reveal the details of the
compact N$_2$H$^+$ emission and to allow a comparison 
with C$^{18}$O.
In Figs.~\ref{cores-b10}-\ref{cores-b218} 
we zoom in towards the regions of bright N$_2$H$^+$ emission
and superpose this emission (in white contours) with 
the emissions of C$^{18}$O and the dust continuum if available (color). 

As discussed before and clearly seen now in 
Figs.~\ref{cores-b10}-\ref{cores-b218},
the C$^{18}$O emission avoids systematically the
regions of brightest N$_2$H$^+$. This was attributed 
to CO depletion, a conclusion that we confirm from
the good match between the N$_2$H$^+$ emission peaks 
and the peaks of the dust continuum. This match 
guarantees that the N$_2$H$^+$ peaks arise from 
true enhancements of the gas column density, and
rules out a possible explanation in terms of
anomalous excitation of N$_2$H$^+$. 
Specially striking is the chain
of four cores shown both in the B211 and B213 panels
(towards the SW in the first one and towards the NE in the second one).
This 0.5~pc-long chain is very prominent both in N$_2$H$^+$ in the 
dust continuum, but is practically invisible in C$^{18}$O.

\subsection{Core selection and physical properties}

As a first step in our analysis of the core population in
the L1495/B213 complex, we make a census of 
N$_2$H$^+$ condensations.
Focusing on the N$_2$H$^+$ emission guarantees an homogeneous
core census, since the continuum maps
(the alternative in a search for cores) do not cover the
full extent of the cloud. It however
results in us missing a small fraction 
of N$_2$H$^+$-poor condensations,
like L1521E in B218, that are known to be at the earliest stages
of evolution \citep{hir02,taf04b}. 

To search for N$_2$H$^+$ condensations, we have first
identified all N$_2$H$^+$ peaks with intensity exceeding
1.2~K~km~s$^{-1}$
and fitted them with 2D gaussians. For that, we have used 
the fitting algorithm of the MOPSIC program
(\url{http://www.iram.es/IRAMES/mainWiki/CookbookMopsic}), which determines
automatically the position, size, and intensity of each emission peak.
In regions where 
several cores overlap or are very close to each other, the fitting 
process has been applied sequentially. In a first step, the brightest peaks 
were fitted and subtracted from the image. Then, a new search for
peaks was carried out in the residual image, fitting additional gaussians
to peaks that still exceeded the intensity threshold. This process was repeated 
until no peaks brighter than the threshold remained.

Table~\ref{tbl_cores} presents the results of our core search. In total, 
19 N$_2$H$^+$
cores were identified in the region. This number likely represents a
lower limit to the number of cores, because if we use an intensity threshold
lower than the chosen 1.2~K~km~s$^{-1}$, the core-finding algorithm 
identifies a  slightly larger number of cores. Visual
inspection of these weaker cores, however, shows that their determination
is uncertain due to the presence of weak and extended N$_2$H$^+$ emission
in several parts of the cloud.
To guarantee that each of our selected cores corresponds to a well-defined
emission peak, we have preferred to
use a conservative criterion based on a relatively high threshold.
Our analysis, therefore, will focus on this 
set of the brightest dense cores of the cloud.

\begin{table}
\caption[]{Dense cores in L1495/B213.\label{tbl_cores}}
\centering
\begin{tabular}{ccccccc} 
\hline 
\noalign{\smallskip}
ID &  $\alpha$(J2000)	&  $\delta$(J2000)  & I(N$_2$H$^+$) & D$^{(1)}$  &  YSO  &  ON$^{(2)}$  \\ 
 & ($^h\;^m\;^s$) & ($^\circ\;'\;''$)  & (K km s$^{-1}$)	 & ($''$) & & \\ 
\hline 
\noalign{\smallskip}
1 & 04 17 43  & 	28 08 03 & 1.6 & 140   & N	 & 	5  \\ 
2 & 04 17 50   & 	27 56 07 & 1.6 & 81    & N	 & 	--  \\ 
3 & 04 17 56   & 	28 12 23 & 1.8 & 149	 & N	 & 	--  \\ 
4 & 04 18 04   & 	28 08 14 & 1.4 & 114	 & N	 & 	8  \\ 
5 & 04 18 04   & 	28 22 34 & 1.2 & 87	 & N	 & 	7 \\ 
6 & 04 18 06   & 	28 05 41  & 2.1 & 157	 & N	 & 	8  \\ 
7 & 04 18 10   & 	27 35 29 & 1.8 & 135	 & N	 & 	9 \\ 
8 & 04 18 34   & 	28 27 37 & 2.8 & 177	 & Y$^{(3)}$ & 11  \\ 
9 & 04 18 41   & 	28 23 22 & 1.2 & 178	 & N	 & 	--  \\ 
10 & 04 19 37   & 	27 15 48 & 2.1 & 139	 & N	 & 	13a  \\ 
11 & 04 19 44  &	27 13 36 & 2.8 & 123	 & Y$^{(4)}$  & 13b \\ 
12 & 04 19 52  & 	27 11 42  & 1.9 & 100	 & N	 & 	-- \\ 
13 & 04 19 59  & 	27 10 30 & 2.0 & 116	 & Y$^{(5)}$ & 14  \\ 
14 & 04 20 15  & 	27 05 59 & 1.1 & 171	 & N & 	--  \\ 
15 & 04 20 59   & 	27 02 29 & 1.8 & 258	 & Y$^{(6)}$ &  16b \\ 
16 & 04 21 21  & 	27 00 09 & 2.6 & 156	 & N	 & 	--  \\ 
17 & 04 27 54  & 	26 17 50 & 2.6 & 137	 & N	 & 	26b \\ 
18 & 04 28 02  & 	26 19 32 & 1.2 & 99	 & Y$^{(7)}$ & 26b \\ 
19 & 04 28 14  & 	26 20 34 & 2.1 & 134	 & N & 	--  \\ 
\hline 
\end{tabular}
\begin{list}{}{}
 \item (1) FWHM of the N$_2$H$^+$ emission (uncorrected for $60''$ beam);
 (2) Core number in the catalog of \citet{oni02}; (3) I04152+2820;
 (4) I04166+2706; (5) I04169+2702; (6) 2MJ04210795; (7) I04248+2612
\end{list}
\end{table}

The results of our core search presented in Table~\ref{tbl_cores}
show that the L1495/B213 condensations have sizes and
N$_2$H$^+$ intensities typical
of the Taurus core population \citep{ben89,cas02}.
Some of these cores are associated with embedded Class 0 and Class I
objects and therefore must have already undergone gravitational
collapse, while others are starless and likely represent
pre-stellar condensations. A number of our N$_2$H$^+$-selected cores have
counterparts in the catalog by \citet{oni02}, which is based on 
H$^{13}$CO$^+$ observations.
Our core list, on the other hand, has very little overlap
with the larger list of extinction-selected cores presented by
\citet{sch10}. As can be seen in the Table 2 of these authors,
these extinction-selected cores have densities on the
order of $10^4$~cm$^{-3}$, and therefore likely represent
a population of condensations at an earlier state of
contraction than our N$_2$H$^+$-selected cores.

\begin{figure}
\begin{center}
\resizebox{\hsize}{!}{\includegraphics{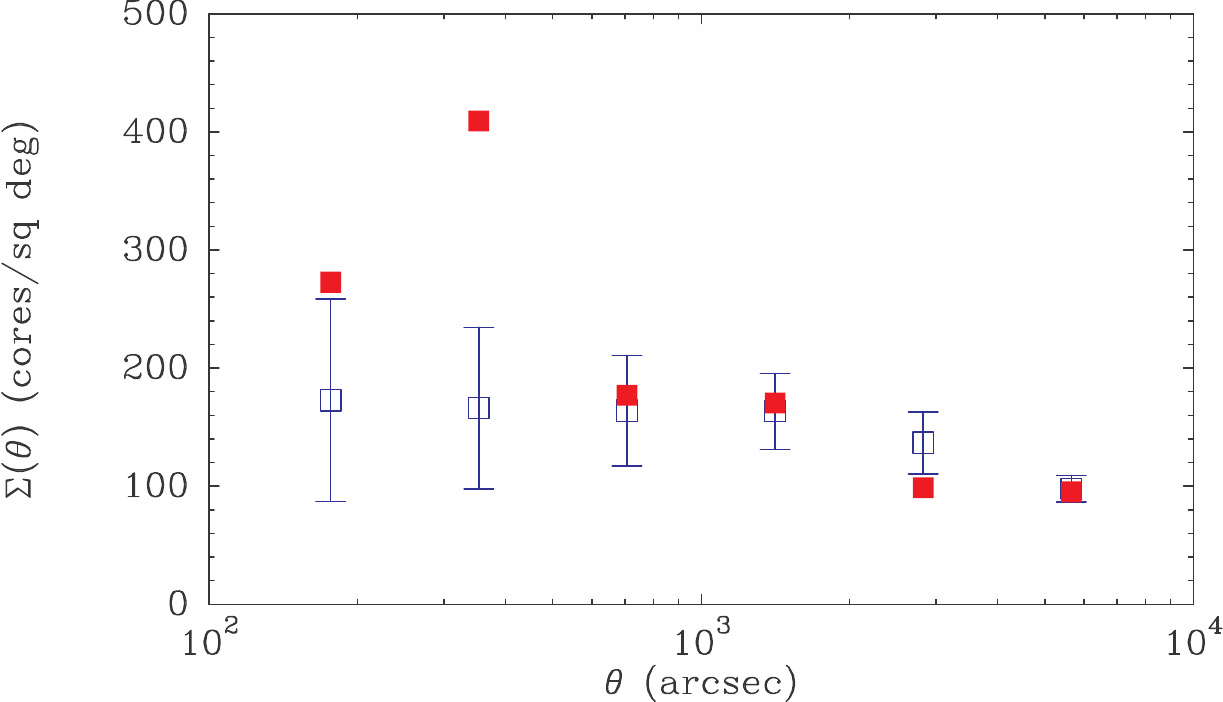}}
\caption{Mean density of companions for the N$_2$H$^+$ cores presented
in Table~\ref{tbl_cores}. The red solid squares represent observations
and the open black squares with error bars represent the prediction from 
a series of 100 Monte Carlo runs (see text).
\label{surf_dens}}
\end{center}
\end{figure}

\subsection{Core clustering}

As we saw in Fig.~\ref{fcrao-large-2}, 
the N$_2$H$^+$ cores in L1495/B213 are not uniformly distributed
over the cloud, but seem
to form small groups of a few objects each.
A similar trend for clustering, but in the lower density gas,
was found by \citet{sch10} using a two-point correlation 
analysis of the distribution of extinction peaks.
In this section, we study the clustering of
the N$_2$H$^+$ cores using the mean surface density of
companions $\Sigma(\theta)$, which is equivalent to the
two-point correlation function but more often used
in studies of stellar clustering (e.g., \citealt{lar95}, see also
\citealt{sim97} for the simple relation between the two). 
Following \citet{sim97}, we have measured for each core
the angular separation to all the other cores in the 
cloud, and we have binned the set of angular separations
in logarithmic intervals with a step equal to a factor of 2.
To convert this number 
of separations into $\Sigma(\theta)$, we have divided the
result
by the area of the bin (which we have assumed rectangular with 
a width of $100''$), and finally we have normalized the
value by 
the total number of cores in the cloud.

The resulting density of core
companions in L1495/B213 is presented in Fig.~\ref{surf_dens}
with solid red squares. To compare it with the expectation 
from a truly uniform distribution of cores, we have 
carried out a set of 100 Monte Carlo simulations. In each simulation,
19 different cores have been created with coordinates
randomly distributed over a rectangle with dimensions 
$13100'' \times 100''$, which is approximately the size of the
C$^{18}$O-emitting region in Fig.~\ref{fcrao-large-1}. Each of
these sets of 19 cores has been treated like the original N$_2$H$^+$ 
data set,
and has been used to calculate a separate mean
surface density of core companions with the same code
used for the data.
The mean of the 100 Monte Carlo experiments is
represented in Fig.~\ref{surf_dens} with open black squares
together with 
error bars indicating the rms value of the dispersion.

As can be seen from the figure, the N$_2$H$^+$  data and the
Monte Carlo
simulation agree within approximately 1-$\sigma$ in the
4 largest bins, indicating that the distribution of cores 
with separations larger than about $700''$ (0.5~pc)
is consistent with being spatially uniform.
At smaller distances, however, the observed distribution of cores
deviates from the prediction of the uniform
model. This is specially striking for
second smallest bin, which is centered at core separations of 
$350''$ (0.25~pc). In this bin, the observed 
density of core companions exceeds the model prediction
by a factor of three, indicating that our sample has three times
more cores with separations of 2-3 core diameters
than expected for a uniform distribution of objects.
Whether a similar or smaller excess of cores occurs at
smaller separations is unclear. The data in the
first bin of Fig.~\ref{surf_dens} presents an excess that is not statistically
significant, but our experiments with the core-finding algorithm
show that this bin is prone to incompleteness
due to our limited angular resolution and to confusion with
extended emission. Higher quality data are
required to analyze the behavior of the core distribution at
distances of 1-2 core diameters.

The excess of cores with separations of about 0.25~pc is
a strong indication that the conditions of core formation 
in L1495/B213 are not randomly distributed over the cloud.
Some regions seem to be specially favorable to
form cores, and therefore give rise to multiple condensations
in close proximity. These regions, however,
cannot extend for much more than about 0.5~pc,
as the mean surface density of core companions 
approximates a uniform distribution at scales
of that size or larger.
As we will see below, this behavior of
the distribution of cores
can be understood if
core formation occurs by the fragmentation of filamentary
structures that have 0.5~pc or so in length, 
provided that the cloud is composed of a large
number of filaments, and that only some of them
have conditions leading to core formation.
The signature of core formation by fragmentation
of 0.5~pc-long structures
seems therefore imprinted in the distribution
of core positions.

\section{Velocity structure: evidence for multiple components}
\label{sect-vel}

\begin{figure*}
\begin{center}
\resizebox{\hsize}{!}{\includegraphics{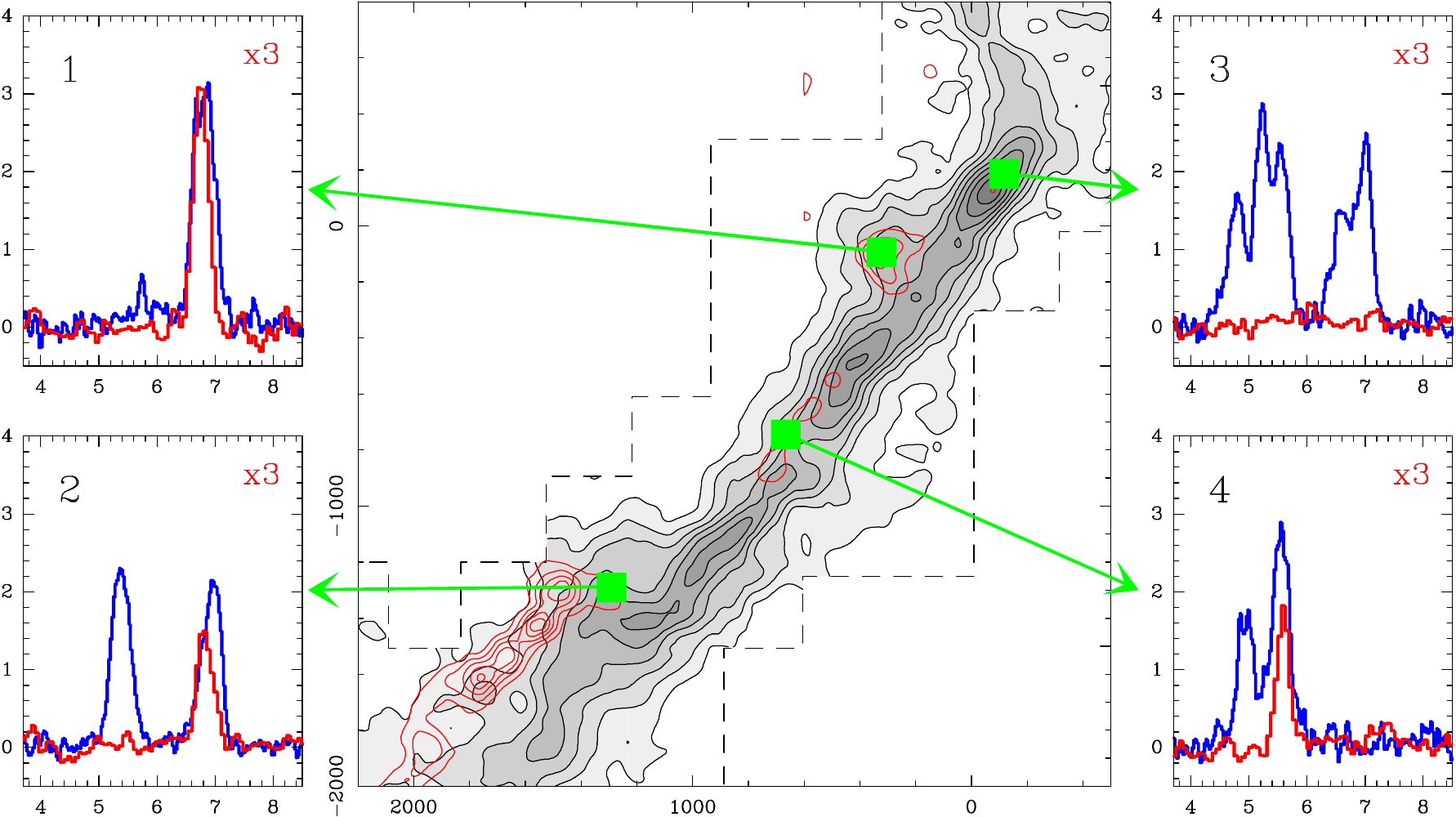}}
\caption{{\em Central panel: } FCRAO integrated intensity maps of the B211-B213 region
in C$^{18}$O(1--0) (grey scale and black contours) and N$_2$H$^+$(1--0)
(red contours). {Left and right panels: } Spectra from selected positions
illustrating the complex velocity structure of the emission. The blue spectra correspond
to C$^{18}$O(2--1) and the red spectra correspond to the isolated 
($F_1 F$ = 01--12) 
component of
N$_2$H$^+$(1--0) (multiplied by 3) observed simultaneously with the IRAM 30m 
telescope. Note the presence of 5 different C$^{18}$O peaks in
panel number 3. 
\label{veloc-comp}}
\end{center}
\end{figure*}

Even taking into account the large-scale fragmentation of the L1495/B213
complex discussed before, the maps of integrated emission and dust continuum
give the impression that the material in L1495/B213 consists of a single filamentary
structure.
The molecular spectra, however, reveal a much more complex picture, and show
that the cloud contains multiple velocity components
that often overlap in projection.
This can be noticed from the inspection of the FCRAO C$^{18}$O(1--0) data, 
but it is better appreciated using a series of high signal-to-noise spectra taken
using the IRAM 30m telescope with the specific goal of characterizing the velocity
structure of the cloud. Some of these spectra are shown in
Fig.~\ref{veloc-comp}, which illustrates the kinematics of the gas in the vicinity of
of the B213 region, the one with the most complex velocity pattern.
Previous observations of this region had noticed the presence of two velocity
components separated by more than 1~km~s$^{-1}$ 
\citep{hei76,cla77,duv86,oni96,li12}.
Our high sensitivity IRAM 30m data reveal now that the kinematics of the region 
is more complex, and
that the number of C$^{18}$O velocity peaks is higher and
changes rapidly with position. 
Fig.~\ref{veloc-comp} shows how each
of two previously known velocity components (near 5.3 and 6.7~km~s$^{-1}$, see panel 2), 
splits in some places
into additional well-separated components, like those illustrated in panel 4.
A more extreme splitting occurs
in the the vicinity of ($-100''$, $200''$), where
the C$^{18}$O spectrum presents 5 separate peaks (panel 3).
Surprisingly, this region with multiple velocity peaks 
corresponds to a single emission maximum in both the continuum and
the FCRAO C$^{18}$O integrated intensity map.

The multiple velocity peaks in the C$^{18}$O spectra of Fig.~\ref{veloc-comp}
seem to originate from true velocity components in the
gas responsible for the emission. 
Self absorption could also create multiple velocity peaks
in a spectrum
by depressing the intensity at intermediate velocities,
but this effect seems to not occur in the C$^{18}$O data. 
The isolated component of N$_2$H$^+$(1--0)
($F_1 F$ = 01--12), which according to hyperfine-structure fits
is optically thin in 86\% of our positions,
always appears if detected at the velocity of one
of the C$^{18}$O components, and not between the two, as
expected in a case of self absorption (e.g., \citealt{leu78}). 
This is illustrated in panels
1, 2, and 4 of Fig.~\ref{veloc-comp}, where the isolated component of
N$_2$H$^+$(1--0) is superposed in red to the multi-component C$^{18}$O
spectrum from the same position. As can be seen, the N$_2$H$^+$ line is associated 
with one of the C$^{18}$O peaks in a way that is inconsistent
with a self-absorption origin of the C$^{18}$O velocity peaks
(the lack of N$_2$H$^+$ in the other
velocity component suggests it has a lower density or a lower N$_2$H$^+$ abundance).
At the extreme position with 5 components in the C$^{18}$O spectrum (panel 3), no 
clear emission is seen in the isolated component of N$_2$H$^+$, but a nearby
position less than $60''$ away presents an N$_2$H$^+$ spectrum with two
weak
components that coincide with the two reddest peaks of the two groups
of C$^{18}$O components. This again argues for self absorption not being
the cause of the multiple peaks seen in the C$^{18}$O spectrum.

The multiplicity of velocity components in the C$^{18}$O spectra
presents a significant challenge for representing and
analyzing the emission.
Velocity structure in a cloud is commonly characterized using channel maps,
in which the emission is integrated in narrow ranges of velocity
that can be selected to highlight particular gas components.
For the L1495/B213 complex, however, this
approach fails due to the large number of components,
their partial overlap in velocity 
(as shown by the spectra in Fig.~\ref{veloc-comp}), and
the fact that each component changes slightly 
its velocity with position.
A better approach to study the velocity field is to search for the
different velocity components directly in the spectra.
Fig.~\ref{veloc-comp} shows that
the velocity components are relatively symmetric when found in isolation
or detected in N$_2$H$^+$, and this suggests that fitting multiple
gaussians to the emission can determine the main properties of 
each component, notably its average velocity and its velocity dispersion.
Fitting
manually an indeterminate number of gaussians to each spectrum in a dataset 
that contains
tens of thousands of spectra is highly impractical, so we have developed a
semi-automatic procedure. 
A number of tests suggested that
the following steps provide a reasonable balance between automation
and supervision.

\begin{enumerate}
\item Divide the mapped area in fields of $150''\times 150''$, 
each containing 25 individual 
spectra.
\item For each field, discard all spectra with no channel
brighter than 3 times the typical rms ($\approx 0.3$~K). If less than 3
spectra are left in a given field, the field is discarded from the analysis.
\item Average all accepted spectra in a field to obtain a high S/N template,
which is inspected visually to determine the number of gaussian components
needed to fit it (in cases of doubt, the individual spectra were inspected).
To be considered real, a component needs to lie at least
0.25~km~s$^{-1}$ (3 channels) apart from the other components
in the spectrum.
\item Fit each spectrum in a field using the number of gaussian components
determined in the previous step. This is done using the
{\tt MINIMIZE} command of {\tt CLASS}. The fit
starts with an initial guess based on the average spectrum of the field, but 
the properties of each gaussian component (intensity, central velocity, and
dispersion) are left unconstrained and fitted automatically. 
\item Discard those fits whose intensity is lower than 3 times the
rms in the spectrum.
\end{enumerate}
With this procedure, it was possible to process reasonably quickly 
the approximately 23,000 spectra of the
FCRAO C$^{18}$O(1--0) map and fit 
a total of more than 11,000 gaussian components with SNR equal to or
larger than 3.
For the N$_2$H$^+$(1--0) data, the analysis was simpler, since no 
FCRAO spectrum showed two velocity components with high enough signal
to noise to deserve a multi-gaussian fit. We therefore applied a 
single-component hyperfine-structure fit (using the {\tt HFS} option
in {\tt CLASS}) and again selected
as real those fits exceeding 3 times the rms in the spectrum. 
Almost 400 N$_2$H$^+$ spectra
were fitted this way\footnote{The fits results for all the components 
with SNR~$\geq$~3 identified in our C$^{18}$O (1--0) and
N$_2$H$^+$ (1--0) FCRAO spectra are
available online at the CDS.}.

\begin{figure}
\begin{center}
\resizebox{\hsize}{!}{\includegraphics{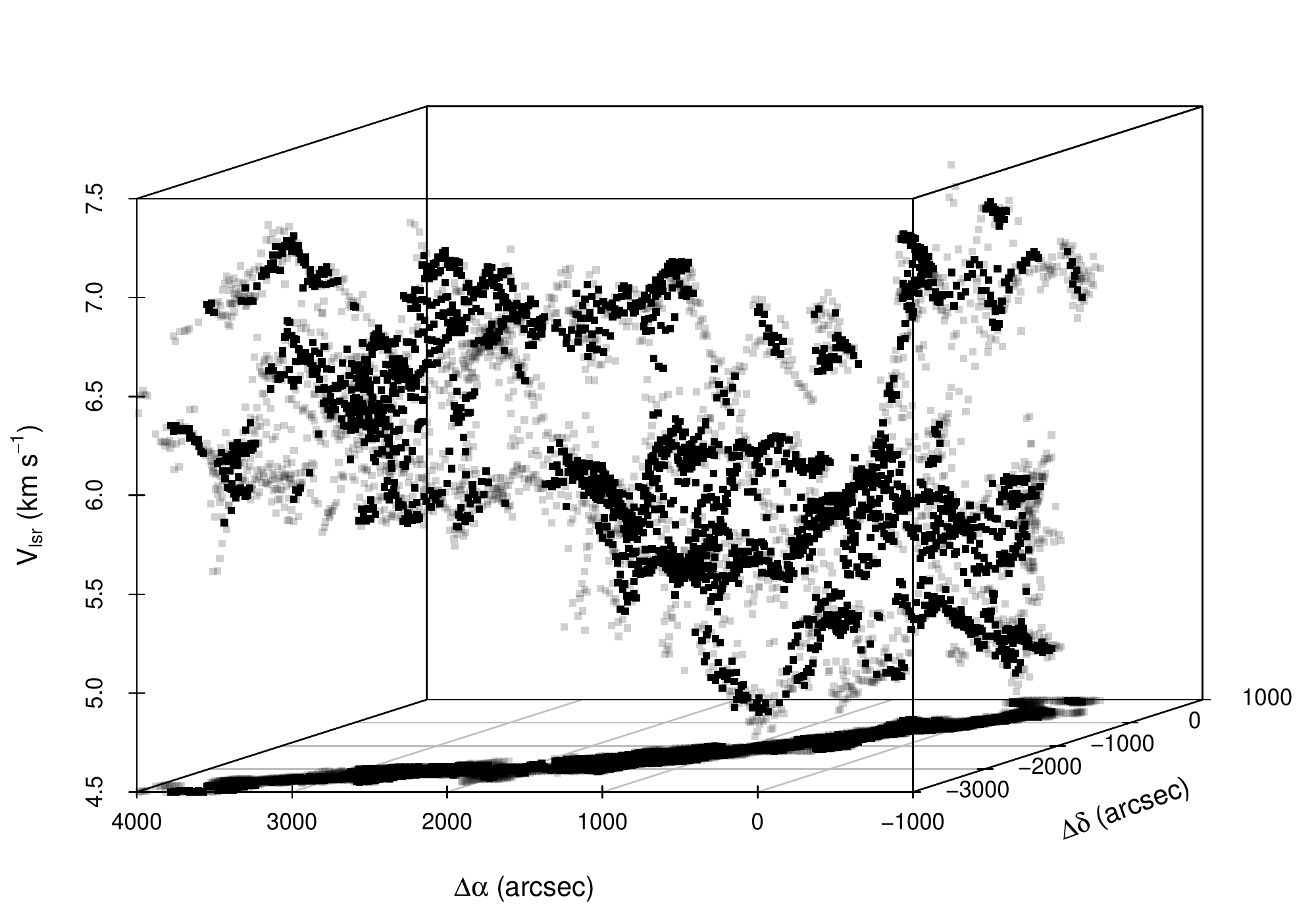}}
\caption{Position-position-velocity cube showing the 
line center velocities derived
from the gaussian fit to the C$^{18}$O(1--0) and N$_2$H$^+$(1--0) 
spectra
towards the B211-B213 region presented in Fig.~\ref{veloc-comp}. Note the
presence of correlated structures. Semi-transparent points have intensities
between 3 and 6 times rms and opaque points have intensities larger than or
equal to 6 rms.
\label{ppv-1}}
\end{center}
\end{figure}

As each gaussian component is characterized by a position in the sky and a 
center velocity (plus a width), the
most convenient way to explore the results of 
the multi-component gaussian fit to the data 
is by plotting the component line center velocities
in a cube of position-position-velocity (PPV)
space. Such PPV space 
represents a subset of the gas phase space
and allows visualizing structures spatially confused in the plane of the
sky but separated in velocity. 
Fig.~\ref{ppv-1} shows one such a PPV cube that covers approximately the
same region presented in the integrated map of 
Fig.~\ref{veloc-comp} and contains the result of our gaussian fitting.
Although undoubtedly complex,
the distribution of points in PPV space shows clear signs of
arising from an organized structure.
Most points cluster in elongated groups that
lie at different "heights" (velocities) and present
smooth and often oscillatory patterns.
These groups are associated with the different velocity components 
seen in the spectra, and their clean separation in the cube shows
how working in PPV space
can help disentangle the complex gas kinematics of the cloud.

The next step in our analysis requires a
procedure to identify and isolate
the different components in the PPV cubes.
Visually exploring those cubes, it is possible to identify a number of
individual groups of points that form coherent structures and
likely correspond to physically distinct cloud components.
It seems however preferable to have an objective algorithm that
not only automatizes the selection, but also provides a quantitative
measurement of the degree of correlation between points in PPV space.

\section{Friends In Velocity (FIVe): an algorithm to 
identify velocity components}
\label{sect-fof}

The problem of identifying velocity structures in a
PPV cube like that of Fig.~\ref{ppv-1} has a number of similarities with
the problem encountered in cosmology of identifying groups of galaxies in
a redshift survey.
Redshift surveys provide catalogs in which each galaxy
is characterized by two coordinates
and a redshift (equivalent to our PPV data), and 
a search for galaxy groups involves identifying 
those sub-sets of 
galaxies that are strongly correlated in both position and redshift.
One of the first and simplest algorithms to 
identify galaxy groups in redshift surveys 
is the so-called friends-of-friends (FoF) method,
initially presented by \citet{huc82} 
and still widely used (e.g., \citealt{ber06}). 
This FoF algorithm starts by selecting a random galaxy in the catalog and 
searching the rest of the catalog
for related galaxies (``friends'') using pre-determined 
thresholds in both position and redshift. Each of the galaxies identified 
as a friend is added to the galaxy group, and 
a new search for friends is made this time centered on the
friend galaxies. The new finds (friends of friends) are added to
the group, and the process is iterated until no more friends
are found, at which point the group is considered complete.
When this happens, a new unassigned galaxy is selected from the catalog and
a new search for friends is started to identify a new galaxy group.
The final product of the algorithm 
is a catalog of galaxy groups and a catalog of isolated
galaxies not belonging to any group (see Fig.~1 in \citealt{huc82}
for a flow chart).

Applying the FoF technique to the line center velocity 
of the C$^{18}$O components
follows the spirit of the cosmology implementation, 
but requires a number of adjustments due to the different nature of the
emitting material. Each galaxy in a redshift survey is a
discrete entity, physically separated from the rest 
(excluding mergers), and therefore well represented by a single point in 
PPV space. A velocity component in a 
C$^{18}$O spectrum, on the other hand, represents
a parcel of a fluid that extends over a large region of space
and has no sharp boundaries. The C$^{18}$O emission, in addition, has been mapped 
with Nyquist sampling and the data points form a dense grid,
in contrast with the sparse sampling of a redshift survey data.
These peculiarities of the C$^{18}$O emission require modifying 
the standard FoF approach, in particular to avoid 
a weakness of the method when applied to
fluids and which we will refer to as ``fragility''.  
This fragility of the FoF method means that 
a single bad point (e.g., from a poor
gaussian fit) can potentially create an artificial
bridge between two components that are otherwise
disconnected in PPV space, and that represent physically
independent entities. While this is not likely to occur in redshift data because
of their sparse nature, it is a real problem when dealing with
the Nyquist-sampled and diffuse C$^{18}$O emission. 
This fragility can lead, for example, to an absurd situation in which 
two distinct velocity components in a spectrum become
connected by the FoF algorithm
because each one is linked through a number of friends 
to a distant and weak point with
an intermediate velocity, either
due to a poor fit or because of a more
complex velocity pattern.

\begin{figure}
\begin{center}
\resizebox{\hsize}{!}{\includegraphics{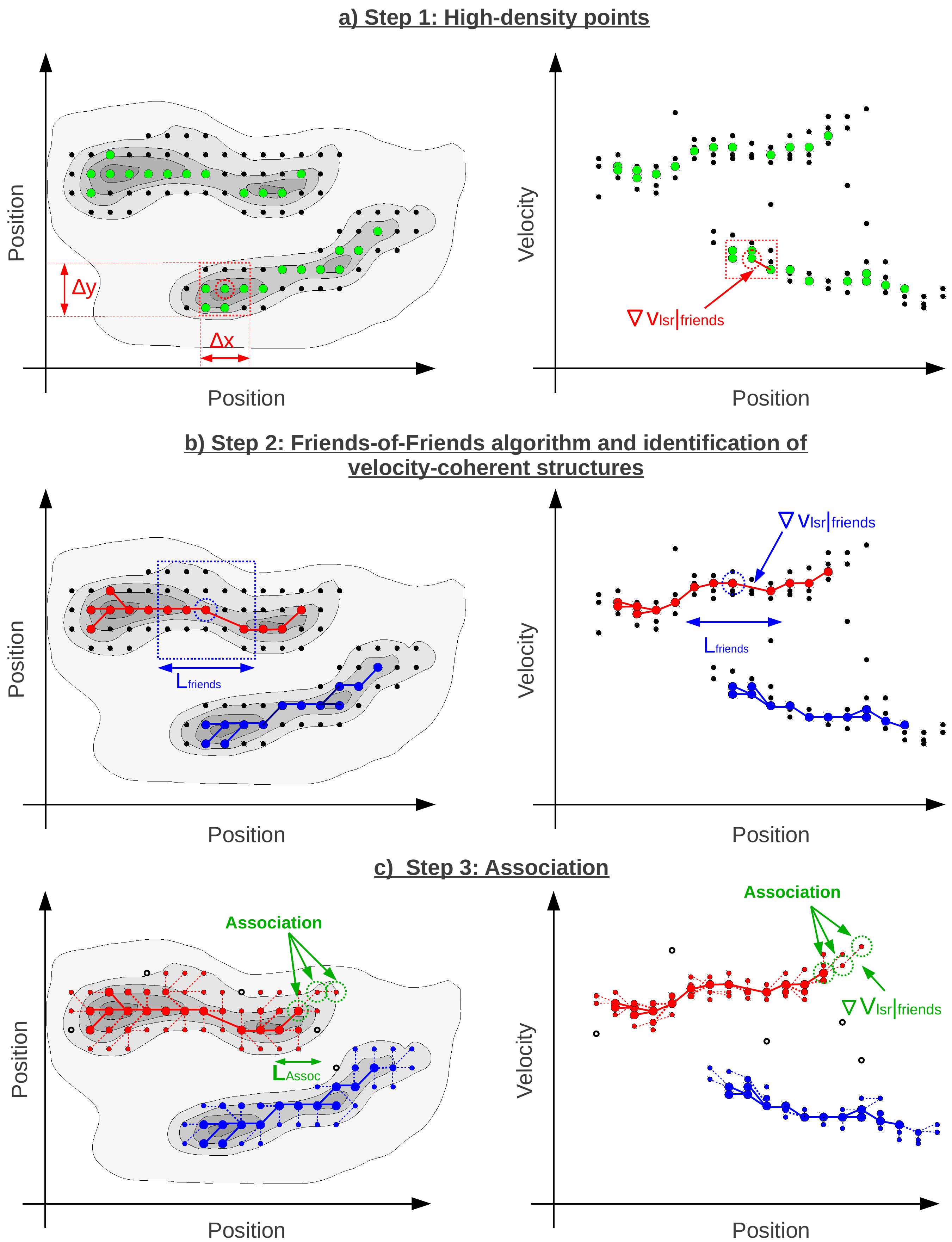}}
\caption{Schematic view of the 3 steps in the modified FoF method 
used to identify
components in PPV space. To simplify the view, two projections of the PPV
cube are presented. The left panels show the PP plane (integrated map) and
the right panels show a projection on a PV plane. 
The black points represent all data with SNR~$\ge 3$. The
green points in the first row of panels are those selected as
significant to use in the first application of the FoF algorithm.
The red and blue colors in the lower panels indicate
positions belonging to each of
the two structures identified by FoF as connected. The
colored lines indicate the individual connection between 
points. Note how points are gradualy assigned to the two 
velocity structures.
The boxes illustrate the regions used to calculate the
number of neighbor points and
the velocity gradients for the FoF algorithm.
\label{fof-cartoon}}
\end{center}
\end{figure}

To mitigate the fragility of the FoF algorithm, we need to 
combine the proximity concept 
of the standard FoF implementation with an additional constraint.
One possibility
is to use the intensity of the emission as an additional criterion
to identify components.
This approach seems the most natural extension of FoF when
dealing with diffuse emission, since intensity itself (which is proportional
to column density) is the criterion already used to define cloud components in
velocity-insensitive datasets like continuum or integrated-intensity maps.
By adding an intensity consideration to the FoF algorithm, our goal
is to make it capable of recognizing that structures like
two long ridges separated by a deep emission valley represent different cloud components,
even if the velocity of the gas in both ridges differs by less
than the FoF-imposed threshold.
Adding intensity weighting to the FoF algorithm is however a complex
task. A number of geometrical and intensity considerations are needed, and
the sophistication of models used to analyze intensity-only continuum data 
from the Herschel Space Observatory testifies the needed level of 
complexity (e.g., \citealt{men12}).
To analyze our data, we have chosen a simplified
method that merges the concepts of FoF and
intensity weighting, and that has been shown to work well in the 
PPV data set produced from our FCRAO observations. The main
idea behind the method is that the emission intensity can be used
to apply the FoF algorithm selectively in a top-down manner, starting 
from the brightest points in the map and working the way downwards 
in sensitivity. A cutoff that selects the brightest 50\% points in the 
dataset is used to define the independent components in PPV space. 
This approach, which is illustrated in Fig.~\ref{fof-cartoon} with an idealized case, 
consists of the following three steps.

\begin{figure}
\begin{center}
\resizebox{\hsize}{!}{\includegraphics{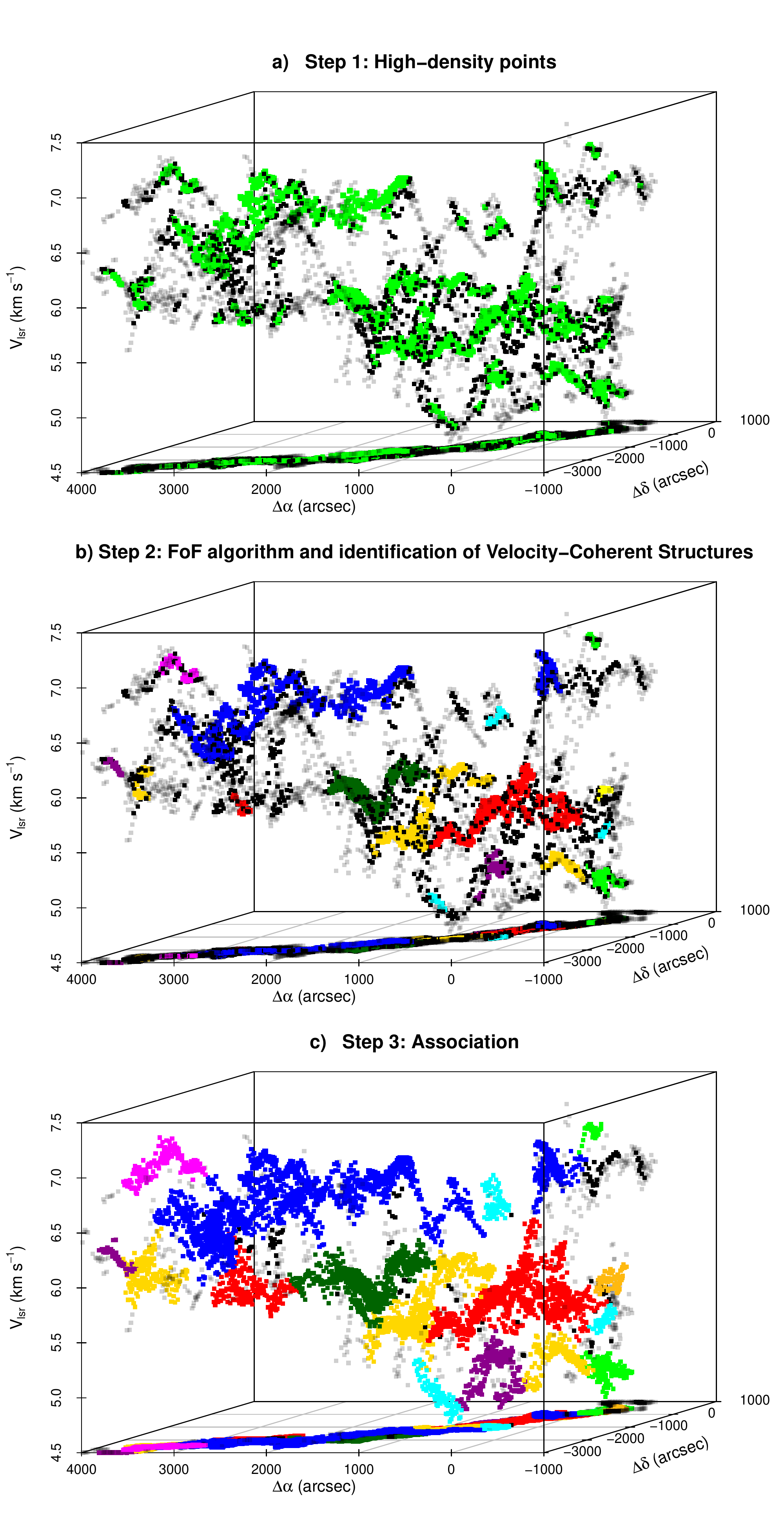}}
\caption{Analysis of the PPV cube of Fig.\ref{ppv-1} using the {\tt FIVe} algorithm. 
{\em Top: } Seed points selected using the threshold values described in the text
(color coded green).
{\em Middle: } Points identified after a friends-of-friends
search that starts with the seed points from the previous step.
The color coding illustrates the different cloud components.
{\em Bottom: } Final assignment of points to the different cloud
components (colored points). The black points represent unassigned points.
\label{fof-result}}
\end{center}
\end{figure}

\begin{enumerate}
\item We first identify the most significant points in the PPV
cube that can serve as seeds for the FoF algorithm.
We do so by finding
positions that either have a SNR $\ge 3$ in N$_2$H$^+$, or
that have both SNR $\ge 6$ in C$^{18}$O\footnote{The intensity of the
C$^{18}$O(1--0) emission in positions with bright N$_2$H$^+$(1--0)
lines (SNR $\ge 3$) has been multiplied by 2 to compensate for the
drop in C$^{18}$O intensity due to freeze out, which is typically
a factor of 2 \citep{taf04a}.}
and more than half of the 8 nearest
neighbors in the $30''$ grid with SNR $\ge 6$ and
a difference in velocity equivalent to
a gradient lower than 3~km~s$^{-1}$~pc$^{-1}$.
(A 3~km~s$^{-1}$~pc$^{-1}$ threshold guarantees that no 
two points in the $30''$ differ in velocity by more than
the typical non-thermal dispersion of the C$^{18}$O(1--0) line, which is 
on the order of the sound speed, or 0.19~km~s$^{-1}$.)
These criteria were found to select the $\approx 35$\% of
points in the sample that lie in
regions of homogeneous properties from which the 
different cloud components can be identified.
\item Using the points selected in the first step, we run a FoF search 
using the C$^{18}$O data with a box of $120''$ and the previous 
3~km~s$^{-1}$~pc$^{-1}$ velocity gradient threshold. The 
choice of the box size, which covers two layers of the $30''$ grid around
the point, allows the algorithm to 
explore larger distances than simply the
next neighbors, and provides an insurance against small-scale drops in the
intensity of the emission that could otherwise fragment the component.
Once the FoF search has been completed, the cloud velocity components are defined
as those groups of friends that are isolated from the rest and that contain
a minimum number of 8 members. 

\item The final step in the analysis consists of assigning the 
points with few neighboring friends and/or
SNR less than 6
to the different cloud components defined in the previous step. 
For this, we relax the previous neighbor and 
SNR thresholds, and test each component point 
for new friends using again the 3~km~s$^{-1}$~pc$^{-1}$ velocity
gradient threshold.
In line with the intensity-weighting scheme discussed before, 
this new search for friends is applied 
sequentially, advancing each cloud component by one step. 
When no new friends are found, the analysis is stopped
and the remaining points (typically $<20$\%) are considered unassigned.
Most of these points are low S/N data or points that belong to what appear to be
separate cloud regions that did not have enough points in step
2 to qualify as a bona-fide component according to our definition.
\end{enumerate}

\begin{figure*}
\begin{center}
\resizebox{15cm}{!}{\includegraphics{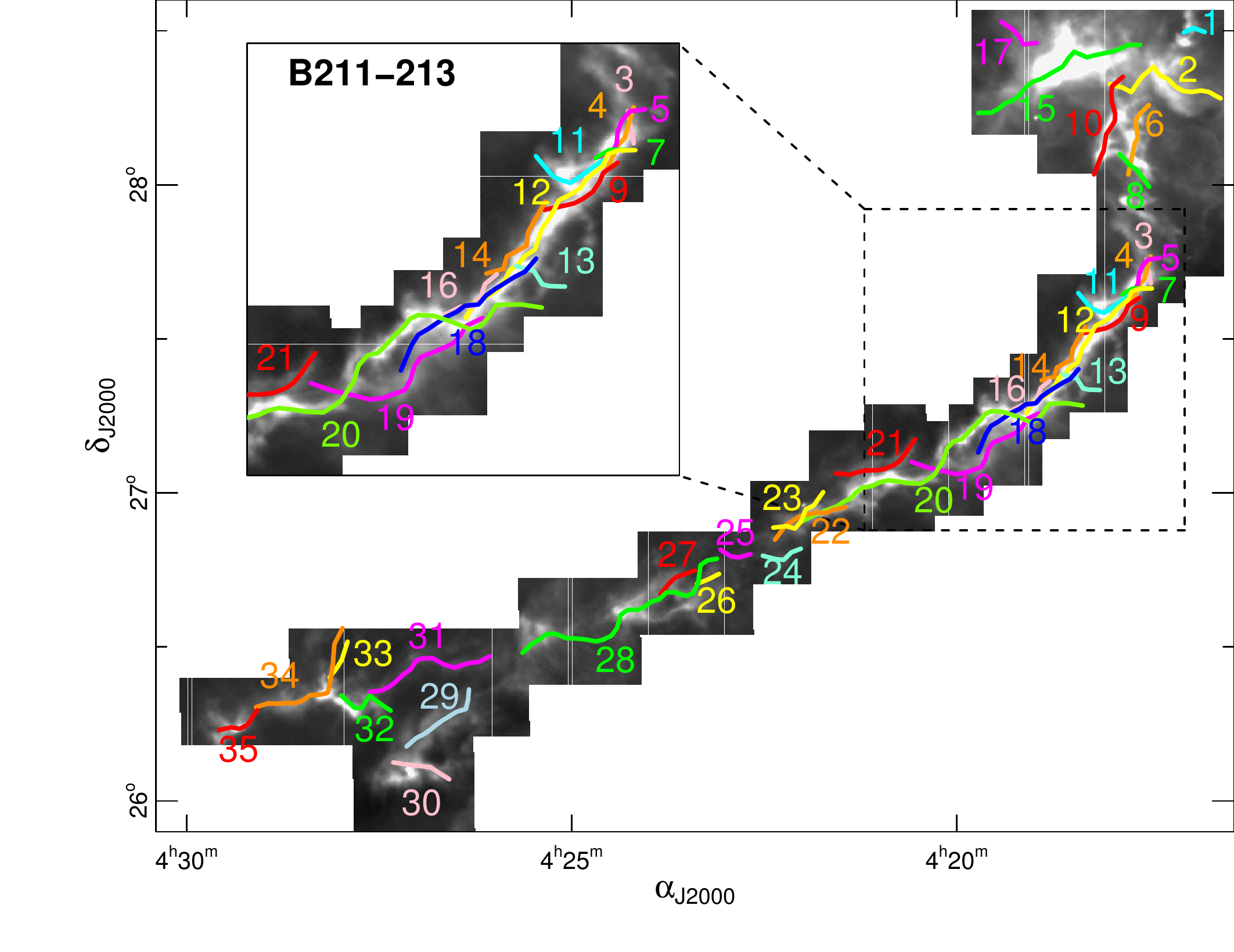}}
\caption{Location of the cloud components of L1495/B213 identified with the
{\tt FIVe} algorithm. Each component is represented by its
central axis and has been color-coded for easier identification. The
background grey-scale image is a SPIRE 250$\mu$m continuum map
from the {\em Gould Belt
Survey } (\citealt{and10}) that has
been blanked out to match the coverage of our FCRAO data.
\label{fil-map}}
\end{center}
\end{figure*}

All the above steps are carried out automatically using 
a dedicated algorithm called
{\tt FIVe} (Friends In Velocity) and programed using
the {\tt R} language ({\url{http://www.r-project.org/}}).
More information on the algorithm and its implementation 
can be found in \citet{hac12}.
A series of tests to determine its success rate and to illustrate
the choice of internal parameters are presented in 
Appendix~\ref{test}.

\section{The nature of the cloud components}

\subsection{Components as filaments}
\label{sect-filam}

\begin{table*}
\caption[]{Main properties of the L1495/B213 cloud components.\label{table_fils}}
\begin{center} 
\scriptsize  
\begin{tabular}{ccccccccccc} 
\hline 
\noalign{\smallskip}
Fil ID  & $\Delta\alpha$\tablefootmark{a}  & $\Delta\delta$\tablefootmark{a}  & Mass  
& L\tablefootmark{b}  & M$_{lin}$\tablefootmark{c}  &  $\langle$V$_{lsr}\rangle$\tablefootmark{d}  
& $\sigma$V$_{lsr}/c_{s}$\tablefootmark{e}  & $\langle\sigma_{NT}\rangle/c_{s}$\tablefootmark{f}  
& $|\nabla V_{lsr}|$\tablefootmark{g}  & Cores\tablefootmark{h} \\  
        & (arcsec)  & (arcsec)  & (M$_\odot$)  & (pc)  & (M$_\odot$~pc$^{-1})$  & (km~s$^{-1}$)  
	&    &    & (km~s$^{-1}$~pc$^{-1}$)  &  \\ 
\hline 
\noalign{\smallskip}
01  &  -654  &  3204  &  1.5  &  0.1  &  11.3  &  5.9  &  0.3  &  0.82  &  1.11  &  --- \\ 
02  &  -420  &  2545  &  30.9  &  1.0  &  32.4  &  7.0  &  1.1  &  0.99  &  0.23  &  --- \\ 
03  &  -181  &  350  &  1.8  &  0.4  &  3.9  &  5.7  &  0.3  &  0.86  &  0.02  &  --- \\ 
04  &  -136  &  228  &  1.2  &  0.2  &  5.4  &  5.3  &  0.3  &  0.81  &  0.62  &  --- \\ 
05  &  -130  &  316  &  4.0  &  0.4  &  9.1  &  4.9  &  0.4  &  0.73  &  0.16  &  --- \\ 
06  &  -99  &  1911  &  5.3  &  0.5  &  11.4  &  6.0  &  1.0  &  1.41  &  1.29  &  1 \\ 
07  &  -50  &  141  &  0.7  &  0.2  &  4.6  &  7.0  &  0.4  &  0.69  &  0.91  &  --- \\ 
08  &  -35  &  1573  &  2.5  &  0.3  &  9.3  &  6.7  &  0.5  &  1.37  &  0.02  &  --- \\ 
09  &  159  &  -237  &  4.3  &  0.5  &  8.7  &  5.0  &  0.7  &  0.83  &  0.18  &  --- \\ 
10  &  183  &  2055  &  15.5  &  0.8  &  20.4  &  6.8  &  0.7  &  1.39  &  0.34  &  3,4,6 \\ 
11  &  305  &  -44  &  6.9  &  0.5  &  14.7  &  6.7  &  0.7  &  1.27  &  0.39  &  7 \\ 
12  &  433  &  -502  &  29.3  &  1.4  &  21.8  &  5.6  &  1.1  &  1.25  &  0.16  &  --- \\ 
13  &  462  &  -981  &  1.2  &  0.3  &  4.3  &  6.5  &  0.5  &  0.84  &  1.10  &  --- \\ 
14  &  633  &  -650  &  5.6  &  0.5  &  10.3  &  5.0  &  1.0  &  0.97  &  0.04  &  --- \\ 
15  &  675  &  2712  &  95.9  &  1.3  &  71.4  &  7.2  &  0.8  &  0.97  &  0.01  &  8,9 \\ 
16  &  993  &  -1156  &  1.6  &  0.4  &  4.5  &  4.8  &  0.8  &  0.81  &  1.28  &  --- \\ 
17  &  1114  &  3092  &  4.3  &  0.3  &  13.1  &  7.2  &  0.5  &  0.84  &  0.38  &  --- \\ 
18  &  1123  &  -1255  &  15.7  &  0.9  &  17.0  &  5.6  &  1.3  &  1.01  &  0.84  &  --- \\ 
19  &  1483  &  -1766  &  18.9  &  1.1  &  17.1  &  5.9  &  0.8  &  1.33  &  0.06  &  --- \\ 
20  &  2259  &  -2015  &  48.3  &  2.5  &  19.7  &  6.6  &  1.2  &  1.15  &  0.11  &  10-16 \\ 
21  &  2637  &  -1960  &  5.7  &  0.6  &  9.1  &  5.9  &  0.7  &  1.07  &  0.27  &  --- \\ 
22  &  3347  &  -2530  &  3.1  &  0.6  &  5.5  &  7.0  &  0.6  &  0.87  &  0.34  &  --- \\ 
23  &  3472  &  -2565  &  4.6  &  0.5  &  9.4  &  6.1  &  0.7  &  1.05  &  0.14  &  --- \\ 
24  &  3630  &  -2999  &  0.7  &  0.3  &  2.7  &  6.3  &  0.4  &  0.58  &  0.92  &  --- \\ 
25  &  4106  &  -2975  &  1.0  &  0.2  &  4.6  &  6.4  &  0.3  &  1.09  &  0.35  &  --- \\ 
26  &  4371  &  -3255  &  1.4  &  0.2  &  7.7  &  7.0  &  0.6  &  0.77  &  1.55  &  --- \\ 
27  &  4740  &  -3248  &  1.3  &  0.3  &  5.0  &  6.7  &  0.4  &  0.76  &  0.82  &  --- \\ 
28  &  5227  &  -3665  &  43.7  &  1.8  &  24.1  &  6.5  &  1.0  &  1.03  &  0.07  &  --- \\ 
29  &  7206  &  -4938  &  9.8  &  0.7  &  14.1  &  6.8  &  0.5  &  0.90  &  0.34  &  --- \\ 
30  &  7369  &  -5465  &  3.9  &  0.4  &  9.1  &  6.5  &  0.5  &  1.17  &  0.35  &  --- \\ 
31  &  7369  &  -4301  &  4.9  &  1.0  &  5.0  &  7.1  &  0.7  &  0.63  &  0.51  &  --- \\ 
32  &  7979  &  -4642  &  8.7  &  0.5  &  17.8  &  7.1  &  0.7  &  0.98  &  0.12  &  17,18 \\ 
33  &  8257  &  -4251  &  0.6  &  0.2  &  3.1  &  6.4  &  0.4  &  0.70  &  1.32  &  --- \\ 
34  &  8578  &  -4493  &  13.3  &  1.0  &  13.4  &  6.9  &  1.1  &  0.90  &  0.65  &  19 \\ 
35  &  9349  &  -4960  &  5.5  &  0.4  &  15.8  &  6.7  &  0.9  &  1.11  &  1.66  &  --- \\ 
\hline 
\end{tabular}
\tablefoot{
\tablefoottext{a} Intensity-weighted centroid measured in offset from the FCRAO map center;
\tablefoottext{b} length of component determined as explained in Appendix~\ref{app-centrd};
\tablefoottext{c} mass per unit length;
\tablefoottext{d} mean velocity of emission;
\tablefoottext{e} dispersion of the line center velocity in sound speed units;
\tablefoottext{f} mean of the nonthermal velocity dispersion in sound speed units;
\tablefoottext{g} mean gradient of the line center velocity;
\tablefoottext{h} associated cores labeled as in Table~\ref{tbl_cores}.
}
\end{center} 
\end{table*}

While Fig.~\ref{fof-cartoon} showed an idealized view of our FoF analysis,
Fig.~\ref{fof-result} presents the true result of applying the {\tt FIVe} 
algorithm to the
the PPV cube of data from the region previously shown in 
Fig.~\ref{ppv-1}. There may be some ambiguity in the 
assignment of the weakest points to the different components, but
the algorithm seems to identify as separate entities
most structures in PPV space that visually appear as distinct components
in the cube. Over the whole cloud, the {\tt FIVe} algorithm identifies 35 different 
components, 17 of them in the region shown in Fig.~\ref{fof-result}. The
main physical properties of these components 
are summarized in Table.~\ref{table_fils} and discussed in the
rest of this section.

The most noticeable property of the L1495/B213 components 
is their filamentary geometry. This can be
seen directly in the PPV cubes, where many of these components 
appear as long and twisted  structures. It is however
better appreciated using standard position-position maps.  
Using these maps, we have determined for each component a
principal axis by connecting with straight lines the 
emission centroids of $90''$-long fragments, as explained
with more detail in Appendix~\ref{app-centrd}.
These principal axes represent the backbones of the components, and
they are shown in Fig.~\ref{fil-map} superposed to the
250~$\mu$m dust continuum Herschel-SPIRE archive from the {\em Gould Belt 
Survey } (\citealt{and10,pal13}).

The most striking aspect of Fig.~\ref{fil-map} is the 
the intricate pattern of 
intertwining filamentary components, especially towards the 
B211-B213 regions shown with detail in the figure inset.
A multiplicity of components was of course expected from the 
multiple velocity peaks seen in the spectra of Fig.~\ref{veloc-comp}.
The map of principal axes, however, shows that the components,
despite their very different velocities, belong to a rather
organized structure.
What at first sight seemed like a single 10~pc-long filamentary
cloud becomes now a complex network of braided 
filaments.

Another remarkable aspect of Fig.~\ref{fil-map} is
how well the principal
axes of the velocity components follow the 250~$\mu$m
dust emission mapped with SPIRE. This is specially noticeable
in the B211-213 region, where most of the components are located.
It suggests that the cloud components traced by the combination
of C$^{18}$O and N$_2$H$^+$ emission correspond to the same 
material traced by the SPIRE observations of the dust emission. Indeed, the mass 
estimated by \citet{pal13} for the B211-213 region matches closely the
mass estimated by our analysis (Table~\ref{tbl_masses}).
The additional kinematic information provided by our line data
show that the continuum-emitting material consists of multiple
velocity components, and this multiplicity, as discussed in the next section,
has important consequences in the estimate of the mass 
per unit length.

\subsection{Statistics of filament properties}
\label{sect-stats}

\begin{figure}
\begin{center}
\resizebox{\hsize}{!}{\includegraphics[angle=90]{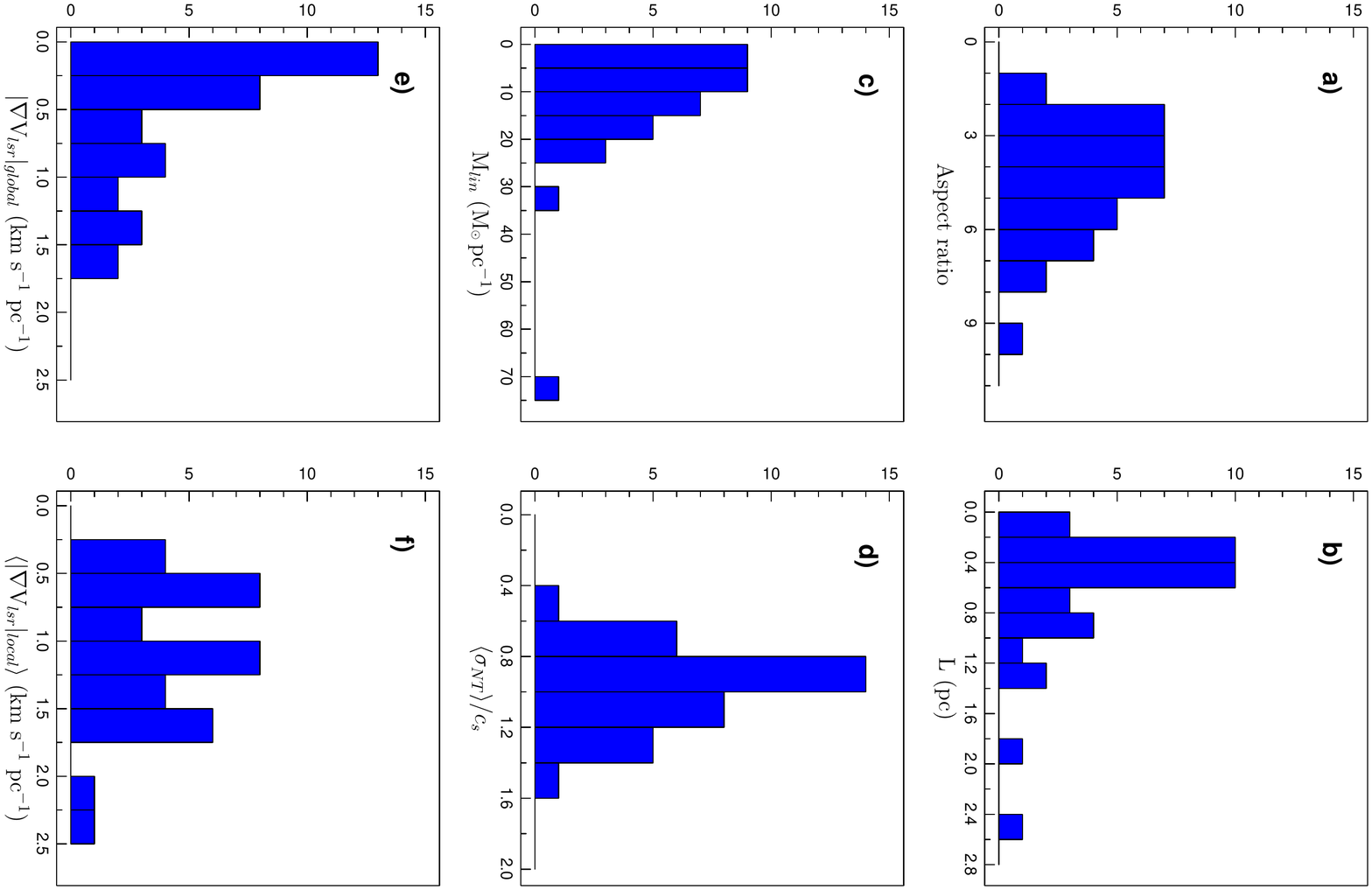}}
\caption{Distribution of parameters in
the 35 cloud components derived using the {\tt FIVe} algorithm.
{\em (a)} Aspect ratio, 
{\em (b)} length,
{\em (c)} mass per unit length,
{\em (d)} non-thermal velocity dispersion,
{\em (e)} global velocity gradients,
{\em (f)} local velocity gradients.
\label{fil-histo}}
\end{center}
\end{figure}

We start the analysis of the filaments 
using a statistical approach.
Fig.~\ref{fil-histo} presents a series of histograms 
showing how the most relevant 
properties are distributed among
the 35 filaments identified by the {\tt FIVe} algorithm.
In agreement with the previous discussion, the histogram of
aspect ratios (top left) shows a distribution significantly
shifted from 1, with 75\% of the components having an aspect
ratio larger than 3. This large value of the aspect ratio provides a
justification for our using the word
``filament'' to refer to most of the observed components.

The statistics of linear sizes is illustrated in the 
top-right histogram of Fig.~\ref{fil-histo}, 
which shows that despite a
significant tail of large 
values, 60\% of the cloud components lie
in a relatively narrow peak between  
0.2-0.6~pc in size\footnote{Some of the longest components may result from
the artificial merging of smaller units.
Indeed, the longest component (number 20) often splits into two separate
components when we run {\tt FIVe} using slightly different
thresholds for the FoF algorithm. Higher sensitivity
data are required to solve this
ambiguity.}.

A related quantity is the mass per unit length, which is presented
in the middle-left histogram (Fig.~\ref{fil-histo}c). This parameter
has been calculated directly from the
C$^{18}$O emission assuming the standard C$^{18}$O abundance of
\citet{fre82}. To compensate for CO depletion, the
contribution from
positions with bright N$_2$H$^+$
emission (SNR $\ge $ 3) has been supplemented with a mass estimated
using the N$_2$H$^+$ emission and assuming standard
excitation and abundance values \citep{cas02,taf04a}.
These N$_2$H$^+$-derived contributions increase the filament mass
typically by 25\% and never reach 50\% of the C$^{18}$O-derived
mass.
As the histogram shows, the 
mean mass per unit length of the components is about 
15~M$_\odot$~pc$^{-1}$.
This value is significantly lower than the 
$\approx 54$~M$_\odot$~pc$^{-1}$ derived by \citet{pal13}
from their Herschel continuum data, but as mentioned 
before, this difference results from the fact that the
continuum analysis associates all the mass to a single
cloud component, while our line analysis shows that 
the mass is distributed among distinct velocity components.

Interestingly, the
derived mean mass per unit length of 15~M$_\odot$~pc$^{-1}$
is very close to the 
equilibrium value for an isothermal cylinder in pressure equilibrium
at 10~K \citep{sto63,ost64}. This property, and both the length and 
aspect ratio discussed before, make the L1495/B213 components very similar
to the velocity-coherent filaments of the nearby L1517 cloud, also 
studied using C$^{18}$O  data. These L1517 filaments
had aspect ratios of approximately 4, typical lengths of
0.5~pc, and mass per unit length in agreement with the 
prediction for a 10~K isothermal cylinder  \citep{hac11}.

A notable difference with the L1517 filaments is the larger velocity
dispersion of the gas in the L1495/B213 components. 
The middle-right histogram in
Fig.~\ref{fil-histo} shows the distribution of the non-thermal
velocity dispersion normalized to the sound speed at 10~K,
as calculated by taking the unweighted mean of the
non thermal linewdiths in the individual gaussian fits
to the spectra
(the dispersion of the individual line center velocities
presents a similar behavior).
While the gas in the L1517 filaments was overwhelmingly 
subsonic (98\% of the points) with a typical
$\sigma_{NT}/c_s = 0.54\pm 0.19$, the gas in 
L1495/B213 presents an approximately sonic dispersion 
of $\sigma_{NT}/c_s = 1.0 \pm 0.2$,
although there is a $\sim 15$\% minority of subsonic, L1517-like filaments.
This larger value of the C$^{18}$O non-thermal motions in 
L1495/B213 does not arise 
only from the action of the embedded protostars (whose effect is noticeable 
but local), and indicates that the gas in most L1495/B213
filaments is more turbulent, or has larger internal velocity gradients,
than the gas in L1517. Even so, the estimated $\sigma_{NT}/c_s$ values
indicate that the
internal motions in the L1495/B213
filaments are at most mildly
transonic, and therefore inconsistent with a possible 
picture of supersonic turbulence.

Probably related to this larger C$^{18}$O dispersion is an also 
larger dispersion of the non-thermal component of 
N$_2$H$^+$. The mean $\sigma_{NT}/c_s$ value for this species is
 $0.61 \pm 0.17$, compared to  $0.36 \pm 0.09$ in L1517.
Apart from this higher velocity dispersion, 
the behavior of the gas in
the L1495/B213 filaments is very similar to L1517,
as we discuss in the following section.

The final two histograms 
present the distribution of velocity gradients 
measured along the long axis of each filament
(bottom panels of Fig.~\ref{fil-histo}).
The left histogram shows the distribution of global velocity gradients, 
which are determined by fitting the full set of
LSR velocity values along the long axis with a single linear gradient.
The right histogram (Fig.~\ref{fil-histo}f)
shows the distribution of local velocity
gradients, which are determined using a similar method,
but this time splitting the velocity data in 0.1~pc fragments 
and then taking the average of the fits.
As can be seen, the mean value of the global gradients
is close to 0.5~km~s$^{-1}$pc$^{-1}$, and the mean value of the local
gradients is about 1~km~s$^{-1}$pc$^{-1}$. 
The larger value of the local gradients results from the presence 
of velocity oscillations inside the components, which average out 
when taking a global gradient. A similar behavior was seen in L1517.

\subsection{Core formation in the L1495/B213 filaments}
\label{sect-core_form}

In section~\ref{sect-cores}  we identified 19 dense cores in the L1495/B213 
complex
based
on their N$_2$H$^+$ emission. Now we study the relation between these dense
cores and the 35 filamentary components identified from the C$^{18}$O
data. Table~\ref{table_fils} shows the assignment of the cores to the different
filaments based on the matching of the N$_2$H$^+$ and C$^{18}$O emission 
in both position and velocity. Cores
2 and 5 were not assigned to any filament because
their surrounding C$^{18}$O emission
did not meet the minimum number of points required
for filament definition in the {\tt FIVe} analysis. 
This lack of association seems a sensitivity problem and not
an indication that these cores are peculiar, since
a {\tt FIVe} analysis using a SNR threshold of 2.5 (instead of the
standard 3 value) identifies C$^{18}$O filaments
associated to these cores (but also suffers from confusion
in different regions).

A notable aspect of the assignment of cores to filaments is that
only 5 out of 19 cores
can be classified as single, in the sense that they belong to filaments
containing no other core (to be conservative, unassigned cores  2 and 5 are treated 
as single). This low number of single cores seems highly improbable, since
there are almost twice as many filaments as cores. A
Monte Carlo simulation shows indeed that the probability of
having 5 or less single cores in the cloud is less than 1\%,
and therefore, that the observed distribution of cores among filaments 
does not arise from a process equivalent to 
throwing randomly 19 cores into 35 independent filaments.
Core formation in L1495/B213, therefore, seems to have occurred 
in a non-random manner, with a small fraction of filaments forming
multiple cores (``fertile'' filaments) and the majority of the
filaments remaining sterile.

The  dichotomy between ``fertile'' and ``sterile'' filaments
provides a simple explanation for
our finding in section~\ref{sect-cores} that there is
an excess of cores with separations around 0.25~pc.
This is so because if
most of the cores in the cloud have formed inside
a small number of fertile filaments,
with each fertile filament generating more than one core, 
the resulting core population is expected to be distributed in small groups,
and each core is expected to have a neighbor at a distance
on the order of a fraction of the filament length. Since the
median filament length is approximately 0.5~pc,
the excess of cores should occur at separations around 0.25~pc,
which is what is shown in Fig.~\ref{surf_dens}.

The fertile/sterile dichotomy of filaments, 
together with the multiplicity of cores
per fertile filament, points to a core formation 
mechanism that depends more on a global property of the 
filament than on a local event in its interior.
To explore this possibility, we have compared
the physical properties of the two filament populations
using the data in Table~\ref{table_fils}.
One noticeable trend is that fertile filaments
have a larger mass per unit length than sterile filaments:
24 vs 10 ~M$_\odot$~pc$^{-1}$ when comparing the means and 
18 vs 9~M$_\odot$~pc$^{-1}$ when comparing the medians.
Although the mass per unit length is undoubtedly an uncertain
parameter due to its dependence on the C$^{18}$O abundance,
part of this uncertainty is mitigated by the relative nature 
of the comparison, and by the fact that all cores
have been analyzed in the same way. 
Interestingly, the data in Table~\ref{table_fils} show that
fertile filaments not only have higher
mass per unit length, but that both their
mean and their median values exceed 16~M$_\odot$~pc$^{-1}$,
which is the equilibrium limit against fragmentation 
for an isothermal cylinder
at 10~K \citep{sto63,ost64}. Sterile filaments, on the other hand, have 
on average a mass per unit length that is below the fragmentation limit
and therefore should be gravitationally stable. 
If this is correct (and it should be tested with
a more accurate mass estimate, preferable from continuum
dust emission), it would suggest that core formation
in the filaments of L1495/B213 simply depends on how much mass the
filament has been able to accumulate. 
Most filaments seem to not reach the fragmentation limit, and therefore
fail to form cores, while a selected few do so and produce multiple 
condensations.

\begin{figure}
\begin{center}
\resizebox{\hsize}{!}{\includegraphics{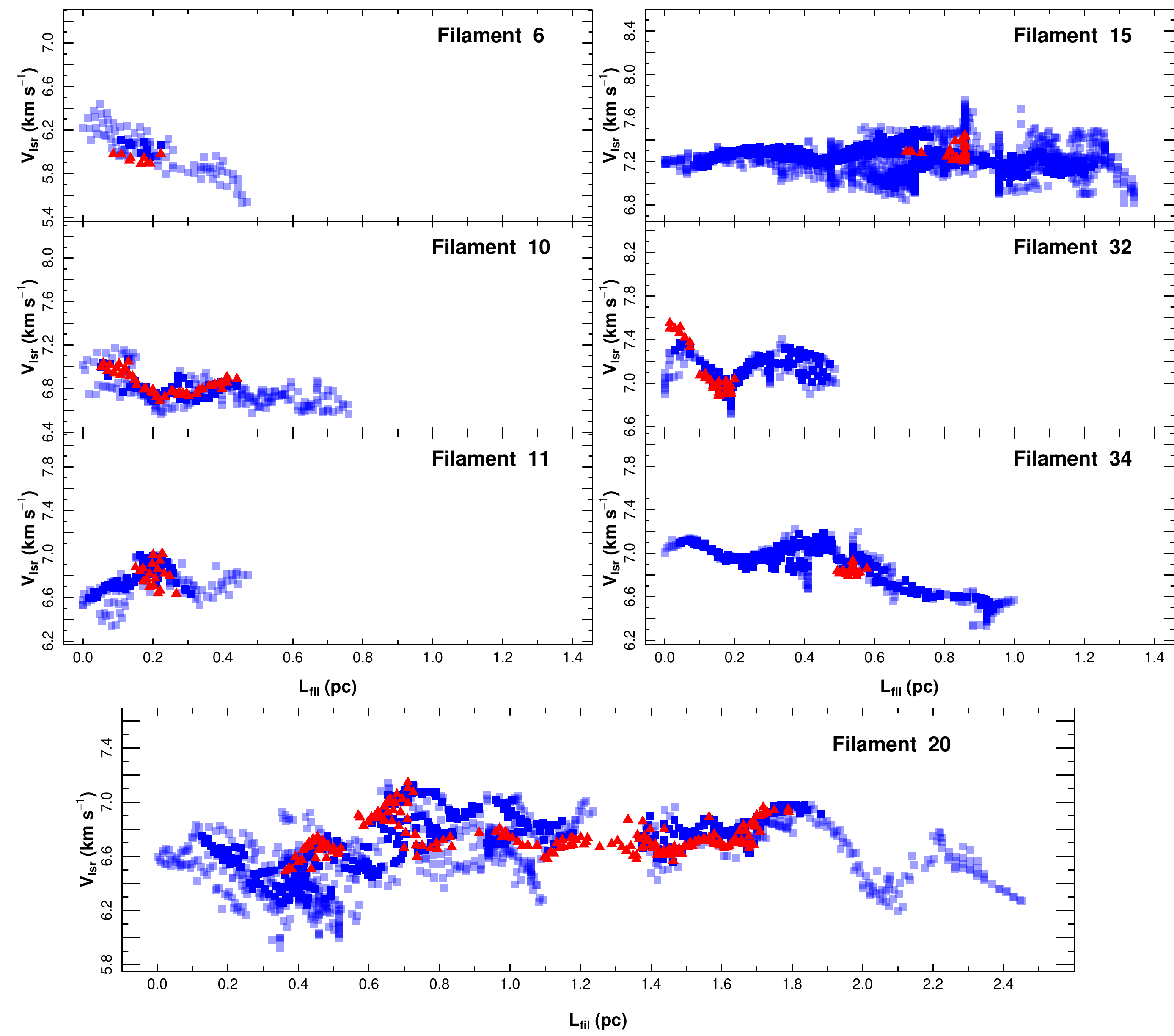}}
\caption{Velocity profiles of filaments with cores. Blue squares are
C$^{18}$O data and red triangles are N$_2$H$^+$ data.
\label{velo-fil}}
\end{center}
\end{figure}

\subsection{Velocity coherence of the filament gas}
\label{sect-vel_coh}

In L1517, \citet{hac11} found that 
the internal velocity field of the cores follows the
large-scale velocity gradients seen in the filaments, suggesting 
that core formation has occurred inside the filaments
with little external interaction (in
contradiction with the expectation
from some turbulent models, see \citealt{kle05}).
To explore this issue in L1495/B213,
we present in Fig.~\ref{velo-fil} velocity profiles
for all filaments containing dense cores.
As can be seen, the gas in the dense cores of L1495/B213
(red triangles) presents large-scale velocity fields
that follow the large-scale gradients of
the lower-density material traced in C$^{18}$O (blue squares),
in good agreement with the L1517 results.
This suggests that the transition from cloud to core
conditions in L1495/B213 has
also proceeded without a
significant change in the gas velocity field.
Core formation, therefore, seems to have resulted
from a quasi-static process that has not decoupled
the kinematics of the dense gas from its
surrounding material.
Internal fragmentation of the filaments,
and not collisions between gas flows, seems therefore 
the core-forming mechanism operating in 
L1495/B213.

Although the velocity field in the cores continues 
the motions of the surrounding material, and 
this suggests that no kinematic changes occur during core formation, 
the non-thermal velocity component of the 
N$_2$H$^+$ lines is on average almost half the 
C$^{18}$O value.
A similar difference was 
found in L1517, and could in principle result
from a lower level of {\em local} random motions in the dense 
core material. This would suggest that turbulence has been
dissipated during core formation.
As argued for L1517, however, 
it is also possible that
the velocity difference between C$^{18}$O and N$_2$H$^+$ 
arises from a combination
of large-scale gas motions and the fact that each molecule 
traces a different column of cloud gas.
The N$_2$H$^+$ emission is very selective of the dense gas 
due to a combination of excitation and chemistry, and
only samples a small fraction of the gas along any line of sight
(as illustrated by the sparse distribution of N$_2$H$^+$ emission 
in the maps of Fig.~\ref{fcrao-large-2}).
The C$^{18}$O emission, on the other hand, is sensitive 
to the extended and lower-density cloud material
due to its close-to-thermal excitation and high abundance,
and it samples a much larger column density of gas than N$_2$H$^+$.
As seen in Fig.~\ref{velo-fil}, all filaments present
large scale velocity gradients in the plane of the sky, and is
therefore likely that similar gradients occur along the line
of sight. Thus, any C$^{18}$O spectrum must contain the
contribution from more parcels of gas moving
at different velocities along the line
of sight than the equivalent spectrum of N$_2$H$^+$, 
and this will undoubtedly cause a larger
non-thermal component.
This effect is confirmed by
the recent numerical simulations of 
core-forming filaments by \citet{smi12}.

The final characteristic of the velocity field that 
we investigate is the 
presence of large-scale organized patterns,
often in the form of
quasi-sinusoidal oscillations 
(see, e.g., filaments 32 and 34 in Fig.~\ref{velo-fil}). 
These patterns, also seen in L1517, suggest that 
the velocity of the gas in the filaments can be correlated
over scales as large as the filament length, or about 0.5~pc.
Such a level of organization in the velocity field seems to deviate 
from the 
expectation for gas belonging to the turbulent cloud regime, which
is characterized by a random pattern that follows a 
linewidth-size relation \citep{lar81}.
Because of that, the L1517 filaments
were referred to as ``velocity coherent,''
extending the term proposed by \citet{goo98} for scales of dense cores.
As Fig.~\ref{velo-fil} shows, the same attribute of coherence seems now
applicable to the L1495/B213 filaments, even if their non-thermal motions 
are close to the sound speed. The L1495/B213 filaments seem to have also decoupled 
from the general turbulent velocity field of the cloud and 
become separate entities with a coherent velocity field.

In L1517, some filaments presented a correlation between
the oscillations of the velocity field and oscillations 
in the filament density, and from that, it was proposed that
the oscillations represent core-forming streaming motions \citep{hac11}.
The origin of the oscillations in L1495/B213
is less clear due to the more complex
velocity patterns and the mix of cores at 
different stages of evolution (i.e, starless and protostellar)
next to each other, which indicates that core-formation in the 
L1495/B213 filaments cannot be treated with the simple perturbation
formalism used in the case of L1517. 
Higher angular resolution observations
of selected filaments are being planned to 
explore the possible connection between velocity oscillations and
filament formation.

\subsection{Filament bundles: a scenario of filament-forming fragmentation}
\label{sect-bundles}

The picture that emerges from the previous sections
is one of hierarchical fragmentation. The L1495/B213 complex
($\sim 10$~pc long) seems to have fragmented first into velocity-coherent filaments
of about 0.5~pc in length. Then, some of these filaments have 
further fragmented into cores ($<0.1$~pc) in an almost quasi-static way. 
Our discussion so far has concentrated on
the second level of fragmentation, the one that breaks up
the velocity-coherent filaments into star-forming dense cores.
This process is better constrained by the data
because both the ``before'' (filament) and ``after'' (core) stages 
can be identified by a different molecular tracer.
In this section
we turn our attention to the first level of fragmentation, the one 
by which the large-scale cloud produces the
velocity-coherent filaments, which
unfortunately is less constrained by the data at hand.

A clue to a possible sequence of filament-forming fragmentation comes 
again from the spectra
of the B211-B213 region shown in Fig.~\ref{veloc-comp}.
As discussed before,
these spectra reveal 
two families of velocity components, one near
5.4~km~s$^{-1}$ and the other near 7.0~km~s$^{-1}$.
Each of these families
further splits into two or three
individual components, which according to our analysis, represent
distinct velocity-coherent filaments. 
Panel 3 in Fig.~\ref{veloc-comp} illustrates how
the velocity differences between the components in each family
are smaller than the differences between the families.
This pattern suggests that
there is some type of hierarchy in the gas velocity field, 
in the sense that the components are grouped into families,
and that the filaments of a given family are more closely connected 
to each other than to filaments of other families.

The idea that filaments are grouped into families is 
reinforced
by the analysis of the physical and chemical properties of the gas.
Filaments with velocities close to 
V$_{\mathrm{LSR}} = 5.4$~km~s$^{-1}$ 
(blue-shifted group of components
in Fig.~\ref{veloc-comp})
are associated with intense emission in C$^{18}$O and SO, 
and at the same time have
very weak N$_2$H$^+$ emission. As discussed in Sect.~\ref{sect-large},
these signatures indicate that the gas has an early-type 
composition, which is expected if the material has condensed 
recently from a more diffuse state.
The group of components 
near V$_{\mathrm{LSR}} = 7.0$~km~s$^{-1}$,
on the other hand,
is associated with bright N$_2$H$^+$ emission and
dense cores containing 
YSOs, which are signatures of more chemically
evolved gas.
The 5.4 and 7.0~km~s$^{-1}$ families of filaments, therefore, 
not only differ in their kinematics, but seem to
be at different stages of evolution.
This behavior strengthens the idea that 
there is a close connection 
between the filaments inside a given family,
while there are noticeable differences between 
the families themselves.

\begin{figure}
\begin{center}
\resizebox{\hsize}{!}{\includegraphics{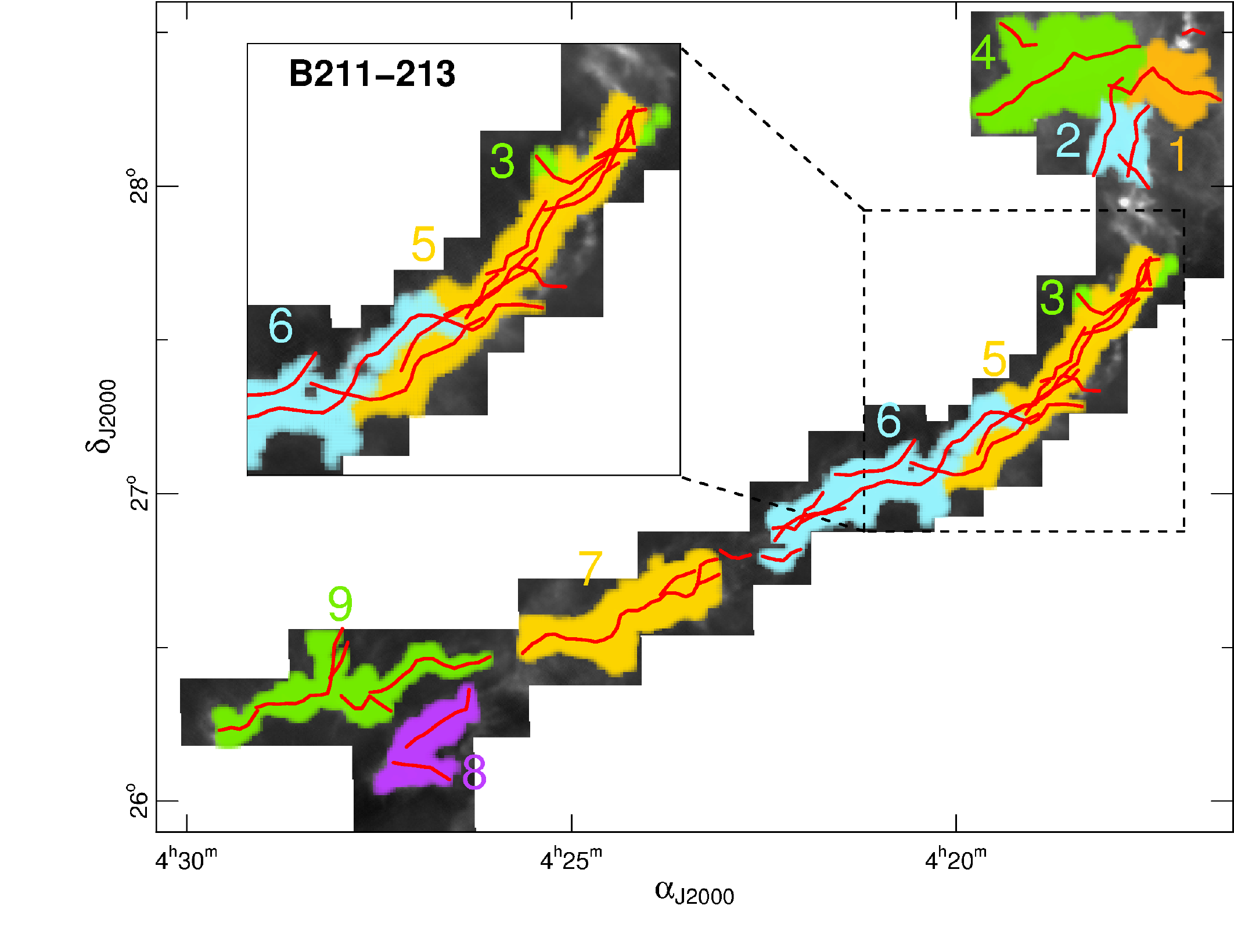}}
\caption{Spatial distribution of the nine bundles of filaments
identified by {\tt FIVE}. The extent of each bundle is 
represented by a colored background, and the axis of
the individual filaments are represented by red lines.
The grey-scale background is the
same SPIRE map shown in Fig.~\ref{fil-map}.
\label{fil-fam}}
\end{center}
\end{figure}

To further explore the possibility that the filaments in 
the L1495/B213 complex are organized into families,
we have used again the {\tt FIVe} algorithm.
This time, we have relaxed its parameters so that it can identify 
structures that are more loosely connected than 
the filaments themselves, but that are still 
noticeably distinct from each other. Motivated by
the differences and similarities between the
5.4 and 7.0~km~s$^{-1}$ families, we have
set the velocity gradient threshold required to
define a ``friend'' to 5~km~s$^{-1}$~pc$^{-1}$.
Also, to favor structures larger than the filaments, we have
required a minimum of 30 points to define a distinct component
and use a box size of $180''$.
The results of this search are shown in
Fig.~\ref{fil-fam}, where the
different families of filaments are identified using different colors.
As can be seen, the
filaments in the L1495/B213 complex seem to belong to a small number of families,
which from now on will be referred to as ``bundles'' because of their 
thread-like appearance. 
(The number of bundles may be smaller than the 9 shown in Fig.~\ref{fil-fam},
since bundle 3 is likely connected to bundle 
6 by low-level emission, \citealt{nar08}.)

While the search for families using the {\tt FIVe}  algorithm 
is strictly based on the kinematic properties of the gas, 
the resulting bundles shown in Fig.~\ref{fil-fam}
reproduce remarkably well the division of the cloud
into regions proposed by Barnard and illustrated in 
the bottom panel of Fig.~\ref{fcrao-large-2}.
In Sect.~\ref{sect-large}, we saw that Barnard's sub-division 
is not simply a morphological one, but that reflects a
separation of the gas into regions that have had
almost-independent contraction histories.
The similarity we find now between these
Barnard regions and the velocity-defined
bundles suggests that the two structures are either
coincident or closely related, and therefore are
likely to have a common origin.
As discussed in Sect.~\ref{sect-large}, the 
different Barnard regions seem to 
have condensed from the diffuse gas with different time scales,
indicating that a large-scale (pc-sized) fragmentation has
occurred early on in the history of the cloud.
If the velocity-defined bundles correspond to these
fragments, bundle-formation must be related to 
the assemblage of the large-scale filamentary cloud,
and therefore must have preceded filament formation. 

A piecewise formation picture of L1495/B213
seems at first to contradict the high degree
of organization suggested by the 
single-filament appearance of the cloud
at large ($\sim 10$~pc) scales
(e.g., \citealt{sch10,pal13}).
This however is not necessarily so
if the L1495/B213 complex was formed by the
convergence of large-scale flows, 
a mechanism favored by a number of 
theoretical considerations and numerical 
simulations (e.g., \citealt{elm93,vaz06,hen08,hei08}).
In this converging-flows scenario, it is
unlikely that gas 
spread over 10~pc linear scales 
can synchronize its convergence without producing 
pc-scale irregularities. For example,
perturbations in the gas velocity on the order of
10\% can easily produce
over the expected 10-20~Myr accumulation time of 
the cloud (e.g., \citealt{ber04}) 
the 1-2~Myr
differences in the convergence time 
required to explain the L1495/B213 data (Sect.~\ref{sect-large}).
Thus, although the 1-2~Myr differences in contraction time
have had a notable influence in the
fragmentation of the cloud into bundles of filaments,
they still represent a small amount from the 
point of view of the full cloud formation process.

While the fragmentation of the cloud gas into families of filaments
seems to arise from local differences in the time-scale of 
gas concentration, 
the internal fragmentation of each family into a small group
of velocity-coherent filaments seems to be intrinsic to
the gas.
Even B211, the youngest region of the cloud, 
consists of multiple velocity-coherent filaments 
despite not having yet
formed stars or cores.
This early appearance of the filaments in the gas suggests that 
their presence results from physical
processes that took place at density regimes lower than those traced
by our C$^{18}$O data (approximately $< 10^4$~cm$^{-3}$).
A number of instabilities are known to occur
at those regimes (e.g., \citealt{hei08}) and may be responsible for
this fragmentation, which has a typical scale length
of 1.5-2~pc. Observations of lower density gas, like
those provided by the more abundant CO isotopes \citep{gol08},
may hold the key to understand the origin of these velocity-coherent filaments.

\section{Summary}

We have studied the 10~pc-long L1495/B213 filamentary complex in Taurus with 
the goal of clarifying the process of dense core formation.
Our data consists of Nyquist-sampled maps 
in lines of C$^{18}$O, N$_2$H$^+$,
and SO, together with partial mapping in the 870~$\mu$m 
and 1200~$\mu$m dust continuum.
From the analysis of these data we have reached the following main
conclusions.

\noindent{\em Large-scale properties of L1495/B213.}
The L1495/B213 complex appears globally as a single large-scale 
filament, but differences across the complex in the YSO population 
and chemical composition suggest that the gas did not 
assemble all at once. Different regions, approximately coinciding
with the condensations originally identified by Barnard
from optical images, seem to have condensed
with time scales that differ by up to 1-2~Myr
(Sect.~\ref{sect-large}).
Embedded in this molecular complex, there is a population of 
at least 19 dense cores, some
of them starless and some of them protostellar. The cores 
are not distributed uniformly, but tend to cluster
with favored separations on the order of 0.25~pc
(Sect.~\ref{sect-cores}).

\noindent{\em Cloud velocity structure.}
The C$^{18}$O spectra often present multiple peaks that indicate the
existence of overlapping velocity components in the gas
(Sect.~\ref{sect-vel}).
We have fitted
multiple gaussians to the C$^{18}$O spectra, and found that 
when the fit parameters are represented
in position-position-velocity (PPV)
space, the velocity components appear 
as coherent structures that are well-separated from each other
(Sect.~\ref{sect-vel}).
To disentangle these components, we
have developed a new algorithm that identifies and extracts automatically
structures in PPV space.
The algorithm, named {\tt FIVe}, uses a friends-of-friends 
approach similar to that often employed to identify clusters of galaxies in 
redshift surveys (Sect.~\ref{sect-fof}).
With its help, we have identified 35 separate components in
L1495/B213. These components are filamentary and 
tend to be aligned with
the axis of the large-scale cloud. They have typical lengths 
of 0.5~pc, internal velocity
dispersions on the order of the sound speed, 
and coherent velocity fields.
They also have mass-per-unit-lengths
close to the fragmentation threshold of an isothermal cylinder at 10~K
(Sect.~\ref{sect-stats}).

\noindent{\em Core formation.}
Despite the large number of filaments in L1495/B213, only a few
of them have formed dense cores, and the
distribution of cores among filaments seems not to be random.
A few ``fertile'' filaments are responsible for most cores
in the cloud, and this suggests that
core formation depends on an intrinsic property of 
each velocity-coherent filament,
such as the mass per unit length (Sect.~\ref{sect-core_form}).
Both fertile and sterile filaments 
present coherent, large-scale oscillations in their
velocity field that suggest that they have decoupled from 
cloud-wide turbulent motions (Sect.~\ref{sect-vel_coh}).
When a filament contains a core, the continuity between the
velocity field of the core gas and that of the surrounding 
less dense material suggests that core formation occurs
with little effect on
the gas kinematics,
in contradiction with the expectation from
models of core formation by gas flow collisions
(Sect.~\ref{sect-vel_coh}).

\noindent{\em Hierarchical fragmentation.}
The 35 filaments of L1495/B213 seem to be grouped
in 9 different families, which we refer to as bundles.
Filaments inside a given bundle have similar kinematics and 
chemical composition, suggesting that they have a 
common physical origin.
This structuring of the gas 
into bundles, filaments, and cores suggests that
fragmentation in the L1495/B213 complex has proceeded in a 
hierarchical manner.
First, the cloud has fragmented into sub-regions (bundles) 
that coincide approximately
with the condensations identified originally by Barnard.
This large-scale fragmentation seems to have resulted from
differences in the local time-scale of gas condensation.
At later times, the sub-regions have fragmented into
velocity-coherent filaments that have typical sizes
of 0.5~pc and velocity dispersions on the order
of the sound speed.
Finally, a small group of fertile velocity-coherent
filaments have accumulated enough mass to fragment
quasi-statically into individual dense cores
(Sect.~\ref{sect-bundles}).

\appendix

\section{Testing and fine-tuning the {\tt FIVe} algorithm}
\label{test}

Our analysis of the L1495/B213 complex has used the
Friends In Velocity ({\tt FIVe}) algorithm
to identify and extract individual gas components 
from the molecular line data. As mentioned
in Sect.~\ref{sect-fof}, this algorithm is based on the 
friends-of-friends method, which has 
a well-documented record of reliability in
the analysis of redshift-survey data
over the last three decades (\citealt{huc82,ber06}). 
Its usage with molecular-line data, 
however, 
exceeds the range of previous
applications, and for this reason, it is necessary 
to extend the algorithm testing using molecular-cloud conditions.
In this appendix, we present a series of tests of the
{\tt FIVe} algorithm under conditions similar to those of the Taurus cloud
to illustrate both the 
successes and limitations of the method.

The analysis of the FCRAO data from L1495/B213 already
provides a number of opportunities to test 
{\tt FIVe}, and different runs of the algorithm 
were carried out to determine how the choice of
internal parameters
(intensity threshold, maximum velocity gradient, etc.)
affects the output results.
It is 
more informative, however, to test {\tt FIVe}
using model data that have
well-defined parameters,
since this allows comparing the {\tt FIVe}-derived 
solution with the original model input.
Experience using the algorithm shows that most 
of its behavior 
can be reproduced with a model that consists of
two filamentary cloud components.
This model is simple enough to allow a systematic 
study of the dependence of the results on the
choice of internal parameters, and thus determine
the conditions under which the 
{\tt FIVe} algorithm succeeds in separating cloud
components, as well as the situations where the algorithm
fails by either merging the components, fragmenting them
artificially, or simply missing them due to low SNR.

The {\tt FIVe} algorithm uses both velocity and
spatial information to identify cloud components.
The role of these two properties is almost orthogonal,
in the sense
that {\tt FIVe} uses different internal parameters
to determine the velocity and spatial structure of the gas.
For this reason, in the following two sections we 
test separately 
the effect of velocity and space, and we do so using
two models: (i) two cloud components that 
overlap spatially but differ in velocity, 
and (ii) two cloud components that have the same
velocity but differ in position.

\subsection{Overlapping components of different velocity}

The case of two components that overlap in space but 
have different central velocities
captures a unique feature of {\tt FIVe}, its ability to
use velocity information to disentangle cloud
structure confused in the plane of the sky. 
Our model for this case assumes
that the cloud is made of
two cylindrical gas components that resemble
the velocity-coherent filaments of L1495/B213.
Each component has a
radial profile of column density parameterized with
a gaussian of $150''$ FWHM, and its
length is $840''$, 
which are typical values
of the C$^{18}$O emission from  L1495/B213
(Sect.~\ref{sect-stats}). The emission from these two components
has been modeled with the {\tt CLASS} program generating
a series of synthetic spectra that have
the same number of channels and velocity resolution as the
FCRAO C$^{18}$O(1--0) data. In each spectrum, the velocity components
are represented by gaussian line profiles that have a FWHM of 
0.48~km~s$^{-1}$ (average C$^{18}$O(1--0) linewidth) and a velocity
separation that is kept as a free parameter to explore 
the level of success or failure of the {\tt FIVe} analysis.
To make the model more realistic, the intensity of the emission,
which is 3~K in the filament axis, has been allowed to vary 
randomly by 20\% (to simulate the
observed non-uniform intensity of the data), and random noise
has been added to each channel of the spectra assuming
an rms level of 0.3~K
(similar to the noise level of our FCRAO observations).

\begin{figure}
\begin{center}
\resizebox{\hsize}{!}{\includegraphics{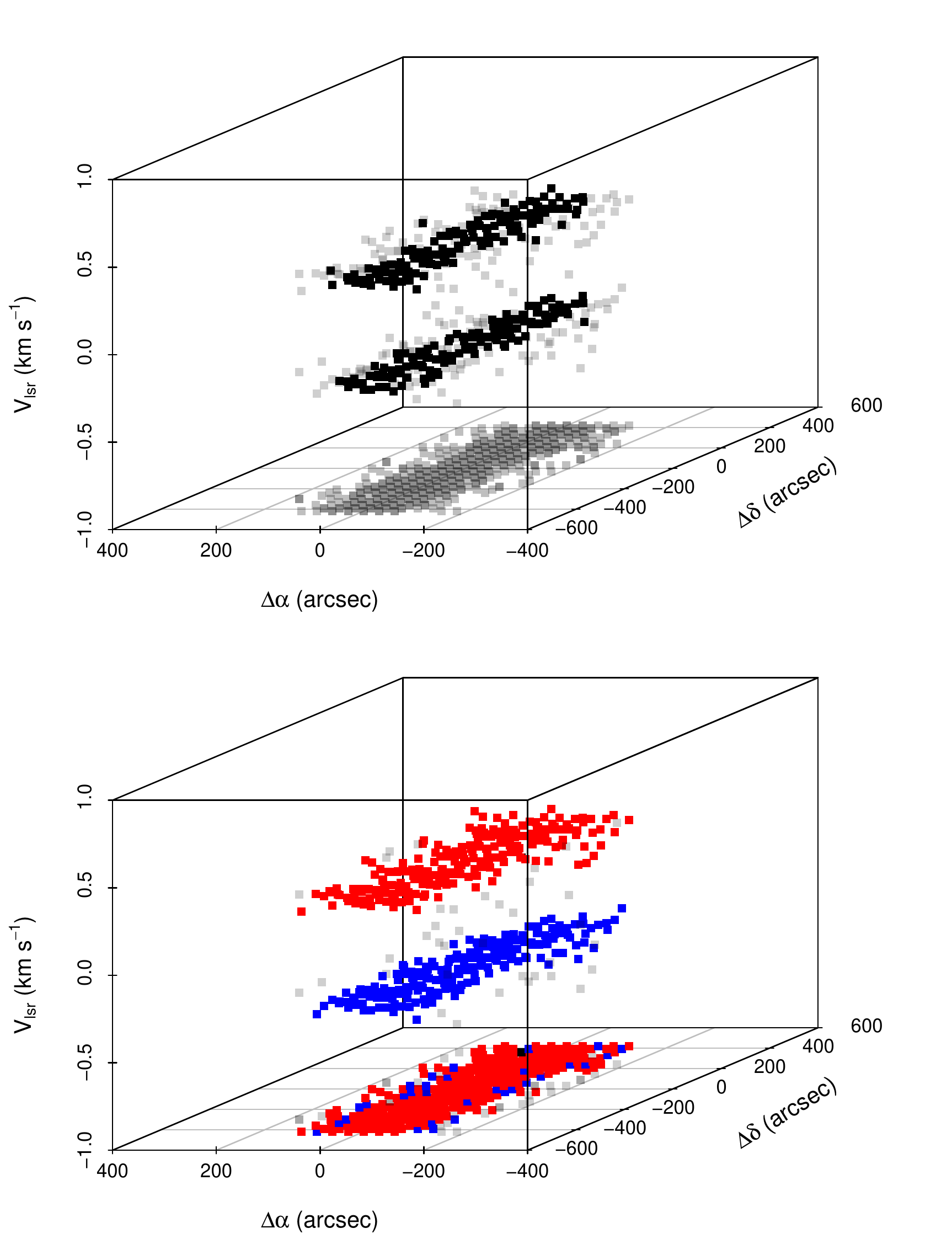}}
\caption{Analysis of a model consisting of
two components that overlap in position 
but are separated in velocity by 1.25 times the line width.
{\em Top: } Position-position-velocity cube with the results
of gaussian fits to the spectra.
{\em Bottom: } Results from the {\tt FIVe} analysis using
the favored choice of internal parameters (see text).
Compare with Fig.~\ref{fof-result}.
\label{ppv-mod1}}
\end{center}
\end{figure}

\begin{figure}
\begin{center}
\resizebox{\hsize}{!}{\includegraphics{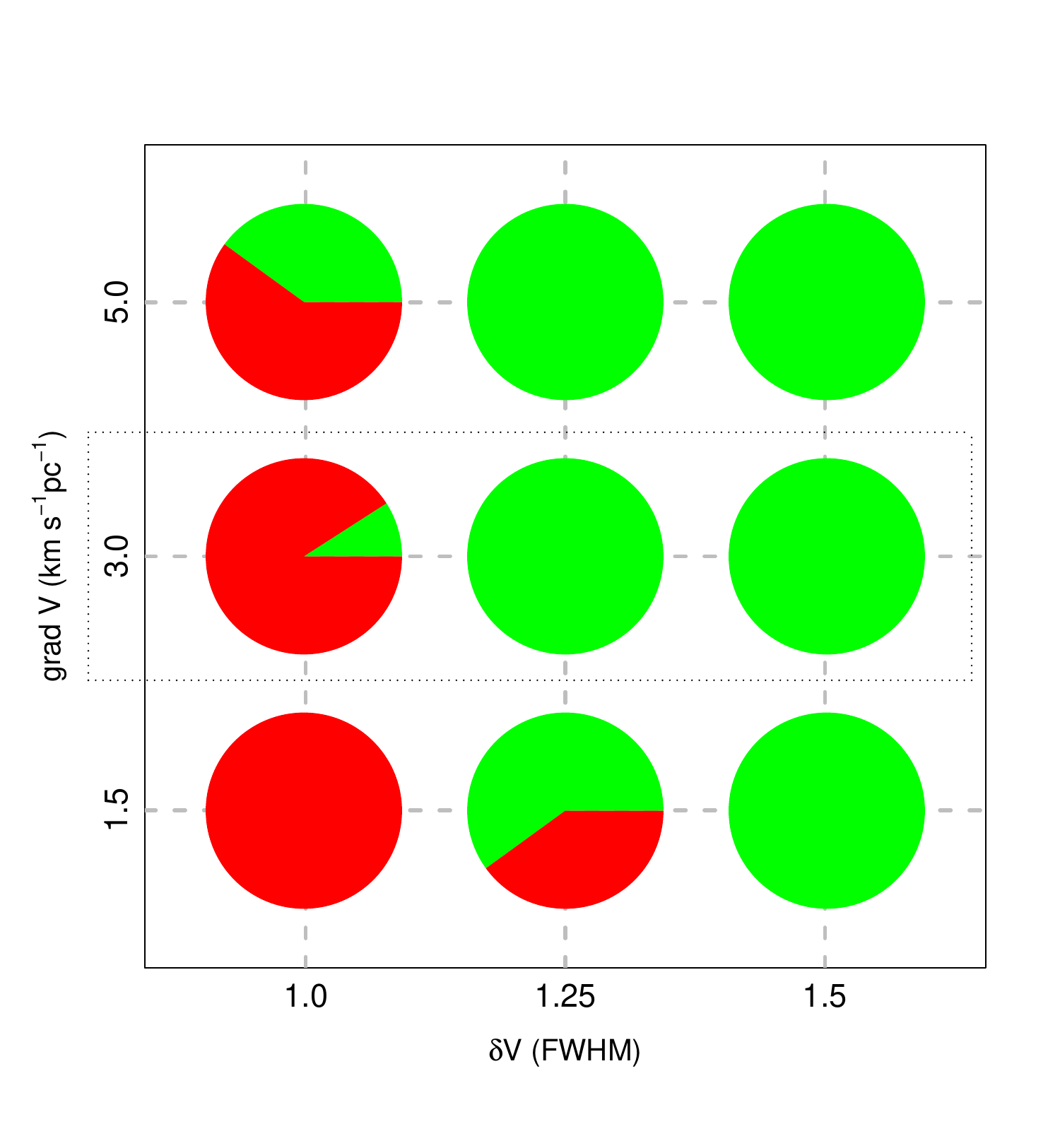}}
\caption{``Traffic-light'' diagram showing the rates of success and failure
of the {\tt FIVe} algorithm when disentangling a model cloud
that consists of two components that overlap in position
but differ in velocity. The diagram represents a matrix of pie charts,
each of them showing 
the fraction of success (green) and failure (red) for 10
independent models with a given velocity difference between 
the components 
(in units of the spectral line FWHM, x-axis) and a
maximum velocity gradient used in the definition of friend
(y-axis). The horizontal box encloses the results for the
favored gradient of 3~km~s$^{-1}$pc$^{-1}$.
(All models use a number of neighbors of 5 and an intensity threshold
of 6 rms, as recommended by the results shown in Fig.~\ref{traffic_2}.)
\label{traffic_1}}
\end{center}
\end{figure}

To simulate the reduction of real data, we
have processed the model
spectra following the same
steps as with the observations.
First, we have fitted the spectra with
two gaussian components, determining 
the line center velocity of the components at each cloud
position. These line center velocities have been used to
generate PPV diagrams like that
shown in Fig.~\ref{ppv-mod1}, and they have been input 
to the {\tt FIVe} algorithm to attempt recovering the
original components.
To explore the ability of {\tt FIVe} recovering
cloud components with different 
amounts of velocity overlap, we have run a series of
models in which the
components differ in velocity by 1.0, 1.25, 1.5 and 2.0 times
the spectral line FWHM.
For each case, we have used different choices of the internal
parameters of {\tt FIVe}, such as the number of required neighbors,
the intensity threshold to select points in the first search
for friends, and the maximum velocity gradient allowed for 
defining friends (Sect.~\ref{sect-fof}). If the {\tt FIVe} algorithm
was able to recover the original two components of the 
model, we have classified the 
test as a success. If not, we have classified the test as
a failure.

\begin{figure*}
\begin{center}
\resizebox{\hsize}{!}{\includegraphics[height=14.5cm]{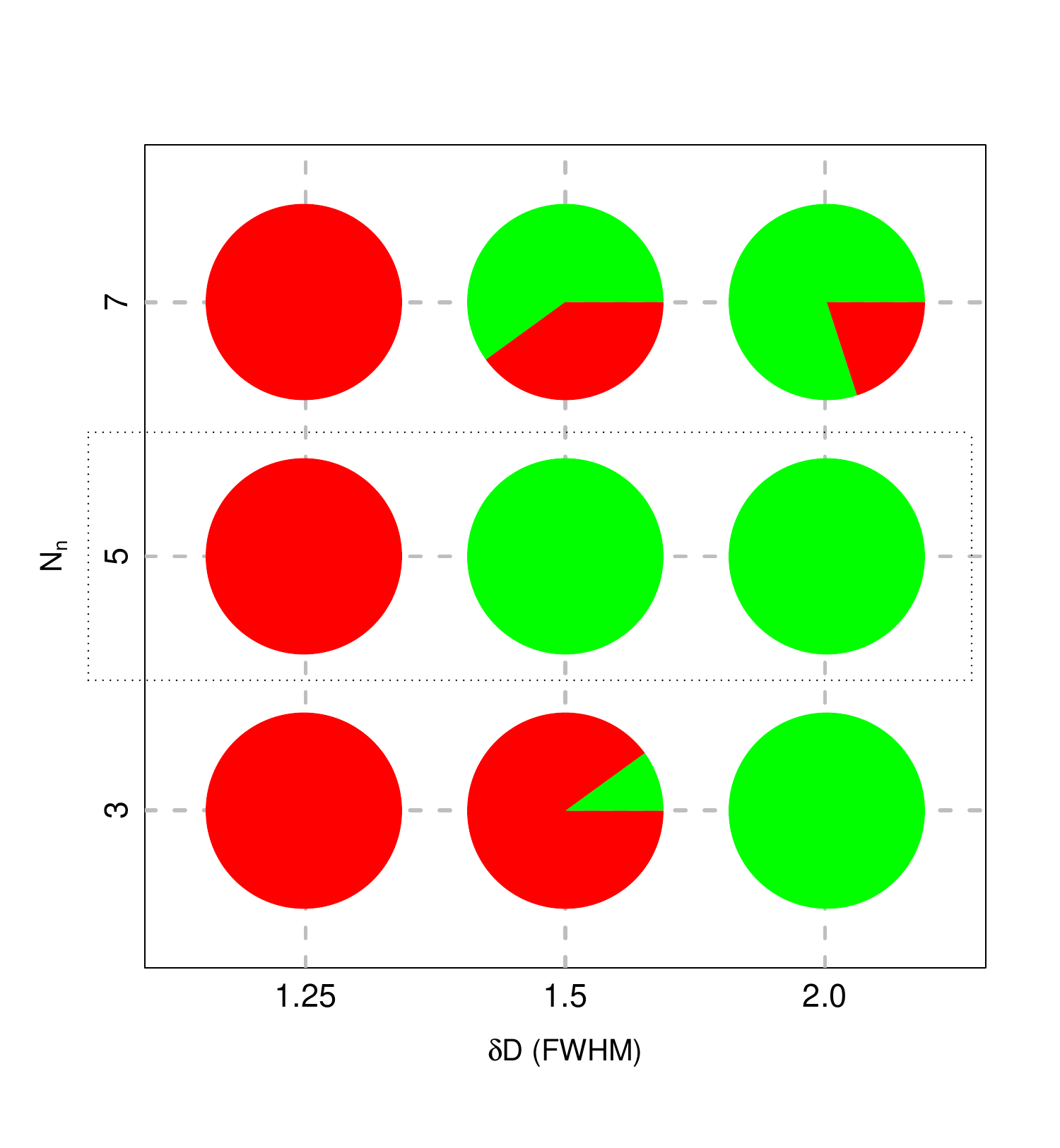}
\includegraphics[height=14.5cm]{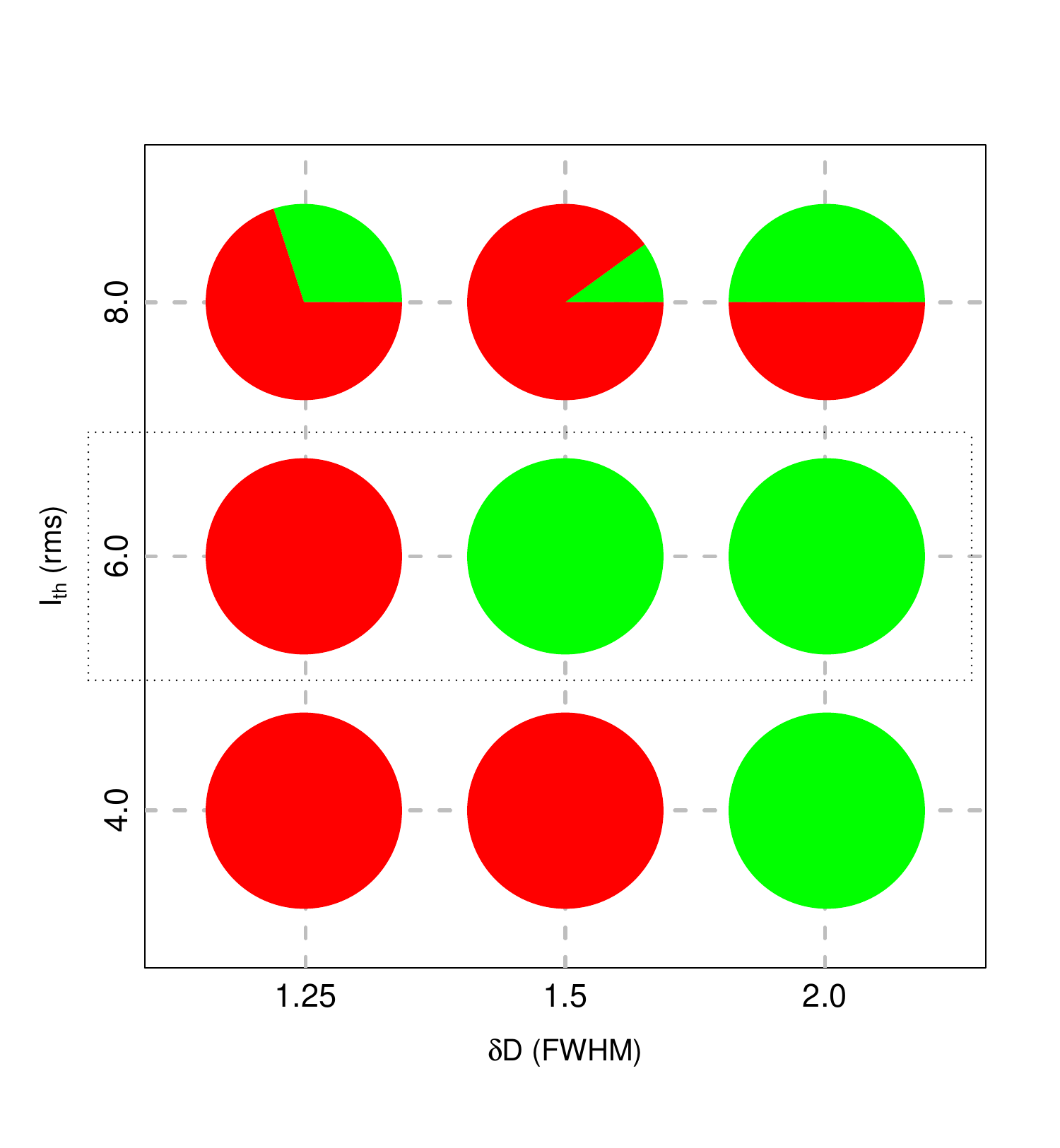}}
\caption{``Traffic-light'' diagrams for a model with two cloud components
that have the same velocity but are separated in the plane of the sky.
{\em Left: } rates of success (green) and failure (red) for a matrix of
separation between components (in units of axial FWHM, x-axis)
and number of neighbors used in the friend definition (y-axis)
{\em Righ: } same as in left panel but for models of variable
intensity threshold (in units of rms).
The horizontal boxes enclose the results for the
choice of parameters favored by our modeling.
(All models assume a gradient of 3~km~s$^{-1}$pc$^{-1}$,
as recommended by the results shown in Fig.~\ref{traffic_1},
and for each panel, the best choice of the other panel is used.)
\label{traffic_2}}
\end{center}
\end{figure*}

Fig.~\ref{traffic_1} presents a summary view of the test results with a
``traffic light'' diagram, which consists
of a matrix of pie charts. Each pie chart 
shows the fraction of successes (in green) and failures
(in red) for a series of 10 independent tests of {\tt FIVe}
that combine 
a given value of the velocity separation between the filaments (in
units of the line FWHM, x-axis) and a value of the 
maximum velocity gradient between the
points used in the definition of friend (y-axis). 
Other internal parameters
of {\tt FIVe}, like the intensity threshold and the number of
points used in the definition of friend, have been kept fixed
at the values used in the analysis of the real data, and
which are 
recommended by the models without velocity structure
discussed in the next section and to which they
are more sensitive.

As can be seen in Fig.~\ref{traffic_1}, for a velocity separation between the
components of 
1.0 FWHM, no choice of the velocity
gradient produces a reasonable ($>50$\%) fraction of
successes. 
This is not surprising, since at this level 
of proximity, the velocity components in the spectrum
start to merge, and the gaussian fitting algorithm often converges to
a wrong result. 
At the other extreme of separations, velocity components
differing by 1.5 FWHM (or more) are disentangled 
successfully by {\tt FIVe} for any reasonable choice of the velocity gradient.
This level of success demonstrates
how {\tt FIVe} can recognize easily cloud components that overlap in
space but are separated in velocity by an amount large enough to
produce a  double-peaked profile in the spectrum.

The critical case in Fig.~\ref{traffic_1} is that of velocity components
separated by 1.25 FWHM (middle column of pie charts). In this case, 
the success rate of {\tt FIVe} depends slightly on the choice of 
the velocity gradient, and is significantly lower ($60$\%) for
the smallest value of 1.5~km~s$^{-1}$~pc$^{-1}$. 
Such small velocity gradient makes {\tt FIVe} prone 
to artificially fragment the components, since
small errors in the velocity determination (due to the 
partial velocity overlap between components) may 
create gaps between sections of the same component
that {\tt FIVe} fails to recognize as connected.
Although not seen in the simple two-component model
of Fig.~\ref{traffic_1},
using a large value for the gradient, like
5~km~s$^{-1}$~pc$^{-1}$, often
leads to the opposite effect, the artificial merging 
of the two components
that are close in velocity if the SNR is low.
As a result, the best choice for the 
maximum velocity gradient used in the friend
determination seems to be close to 
3.0~km~s$^{-1}$~pc$^{-1}$, which is the value
we have used in the analysis of the real
L1495/B213 data.

\subsection{Non overlapping components of the same velocity}

Our second set of models tests the ability of {\tt FIVe} 
to separate cloud structure based on spatial information. 
These models assume that the
cloud consists of two components that have the same radial
velocity but are separated in the plane of the sky by an amount that 
we treat as a free parameter. These models are complementary
to those of the previous section, and 
can be used to constrain a different set of
internal parameters of  {\tt FIVe}, those that are related to spatial
information, like the number of points required to define a friend
and the intensity threshold required to consider points in the search
for friends. 

For this second set of tests, we have followed
the same steps described in the previous section. First,
we have generated a series of model spectra using
the {\tt CLASS} program.
For this, we have assumed that each individual cloud component
has the same 
physical dimensions, velocity structure, emerging intensity,
and noise level as the components in the previous section.
Then, we have fitted gaussians to the spectra, although 
this time, we have used single gaussians because the two 
cloud components do not overlap
in space and the spectra consist of single peaks.
Finally, we have input the fit results into {\tt FIVe}, and we
have compared 
the number and properties of the cloud components
found by the algorithm with those originally put in the model.
As before, we have classified the result as a success if {\tt FIVe}
recovers the original structure. If not, we have classified it 
as a failure.

Fig.~\ref{traffic_2} summarizes the results of our second set of models
using again ``traffic light'' diagrams.
As mentioned before, the models in this section 
explore mainly two parameters of {\tt FIVe},
the number of points used in the definition of
friends and the 
intensity threshold used for the selection of points.
For this reason, Fig.~\ref{traffic_2} consists of two panels,
each one showing the success rate of one of the parameters
when disentangling two components that are separated by a
distance value given
in terms of the filament width (FWHM).

As can be seen, no choice of parameters allows {\tt FIVe} to 
disentangle filaments that are closer than 1.25 times the 
spatial width
or less. 
This limiting value is larger than the 1.0 times the velocity linewidth
found in the previous section,
indicating that {\tt FIVe} is not as sensitive when using spatial 
information as it is when using velocity data.
The (slight) asymmetry between space and velocity 
likely arises from the simpler treatment that
{\tt FIVe} makes of the spatial information, which in turn
results from the more complex nature of this parameter, that 
can vary in two dimensions, while the radial velocity 
has only one dimension to change.

Another result seen in Fig.~\ref{traffic_2}  is the success of
{\tt FIVe} to disentangle components that are separated by
2.0 FWHM (or more) in distance unless some of the internal parameters
of the algorithm
are set to extreme values (e.g., intensity threshold larger
than 8 rms). This level of success indicates that 
{\tt FIVe}  can reconstruct correctly cloud structure that is 
well-separated in the plane of the sky, and that it only
tends to fail when the two cloud components overlap in space.

Fig.~\ref{traffic_2} also indicates that the 
success rate of {\tt FIVe} can be maximized by choosing
appropriate values for both the number of neighbors
and the intensity threshold
used in the friend definition.
As can be seen, a choice of 5 neighbors and 
a threshold equal to 6 times the intensity rms provides a
high success rate even in the case of components
separated by 1.5 FWHM. For this reason, the previous choice of 
values was used in the analysis of the L1495/B213 data.
It is important to stress, however, that this parameter choice
depends on the physical properties of the gas
and the sampling and sensitivity of the observations.
Different cloud conditions or a different type of observation
may require using different parameter choices. 

To finish our analysis, we study how the {\tt FIVe} algorithm
fails in the limiting case of components separated by 1.5 FWHM
when its parameters have not been optimized. This is instructive
because it shows the compromise 
that is often needed to determine the optimal value of a 
given parameter. 
As Fig.~\ref{traffic_2} shows, 
in the case of a variable number of neighbors (left panel),
a choice of 5 neighbors produces a higher rate
of success than a choice of 3 or 7 neighbors.
This is because a choice of a too few neighbors, like 3, can 
lead to the erroneous merging
of the components, since it becomes possible that a statistical
fluctuation due to noise or variations in intensity affects
a few positions that can find enough neighbors to 
create an artificial bridge between separated components.
On the other hand, a choice of too many neighbors, like 7, 
can lead to the erroneous splitting of a single component.
This is because a requirement of 7 neighbors for the definition
of a friend leaves too little room for possible noise effects.
A statistical fluctuation
that eliminates several of the 8 possible nearest neighbors
breaks the component in two and makes the algorithm fail.

A similar situation occurs with the choice of the intensity threshold, as
shown in the right panel of Fig.~\ref{traffic_2}. Choosing too low a value 
(like 4 rms) allows the low-intensity tails of the gaussian 
distributions to merge, and this makes the algorithm fail to separate 
the two cloud components. 
Choosing too large a threshold (like 8 rms), on the other hand, 
makes the algorithm sensitive to the loss of points in
statistical fluctuations, and this leads again to artificial
fragmentation. The optimal value of choice
(threshold of 6 rms), therefore, represents a compromise between 
the two opposite trends of merging and fragmentation, and
makes it possible to disentangle cloud components
separated by 1.5 FWHM.

\subsection{Summary: successes and limitations of {\tt FIVe}} 

While our two-component cloud can only be considered a toy model, its analysis
with {\tt FIVe} presents remarkable similarities with the analysis of the real 
L1495/B213 data. This is likely a combination of the realistic
emission parameters of the model and the fact that most problems 
analyzing of real data arise when dealing with
pairs of components, especially when trying to separate them
if they are close in velocity or position.
Under these conditions, the previous tests  
show that  {\tt FIVe} can recognize components
with an accuracy that approximates that of a human 
observer who has the patience of inspecting the individual spectra
and who looks for connections between spectral features at different
positions. Reproducing such a human behavior was in fact
the main goal when designing {\tt FIVe}, given the large size of the 
L1495/B213 dataset and the need to automatize the 
velocity analysis.

The success of {\tt FIVe} is remarkable in view of its
simple-minded analysis, but is accompanied by 
important 
intrinsic limitations.  As we have seen, {\tt FIVe}
cannot 
disentangle 
components that are blended either in position or velocity
by amounts comparable to their spatial or
velocity width, 
even if the blending occurs over a limited region in space.
Improving on that limit may require a significantly
different type of analysis. Like similar algorithms,
{\tt FIVe} has only a local view of the emission,
and cannot take advantage of the information 
provided by the data on global scales 
to separate so-far unresolvable components.
Adding such a global view to {\tt FIVe} not only requires a more
sophisticated numerical scheme to combine local and
global patterns, but an a-priori
understanding of the physical nature of the components themselves.
This understanding, unfortunately, 
still eludes us today.

\section{Principal axis determination of the velocity components}
\label{app-centrd}

\begin{figure}
\begin{center}
\resizebox{\hsize}{!}{\includegraphics{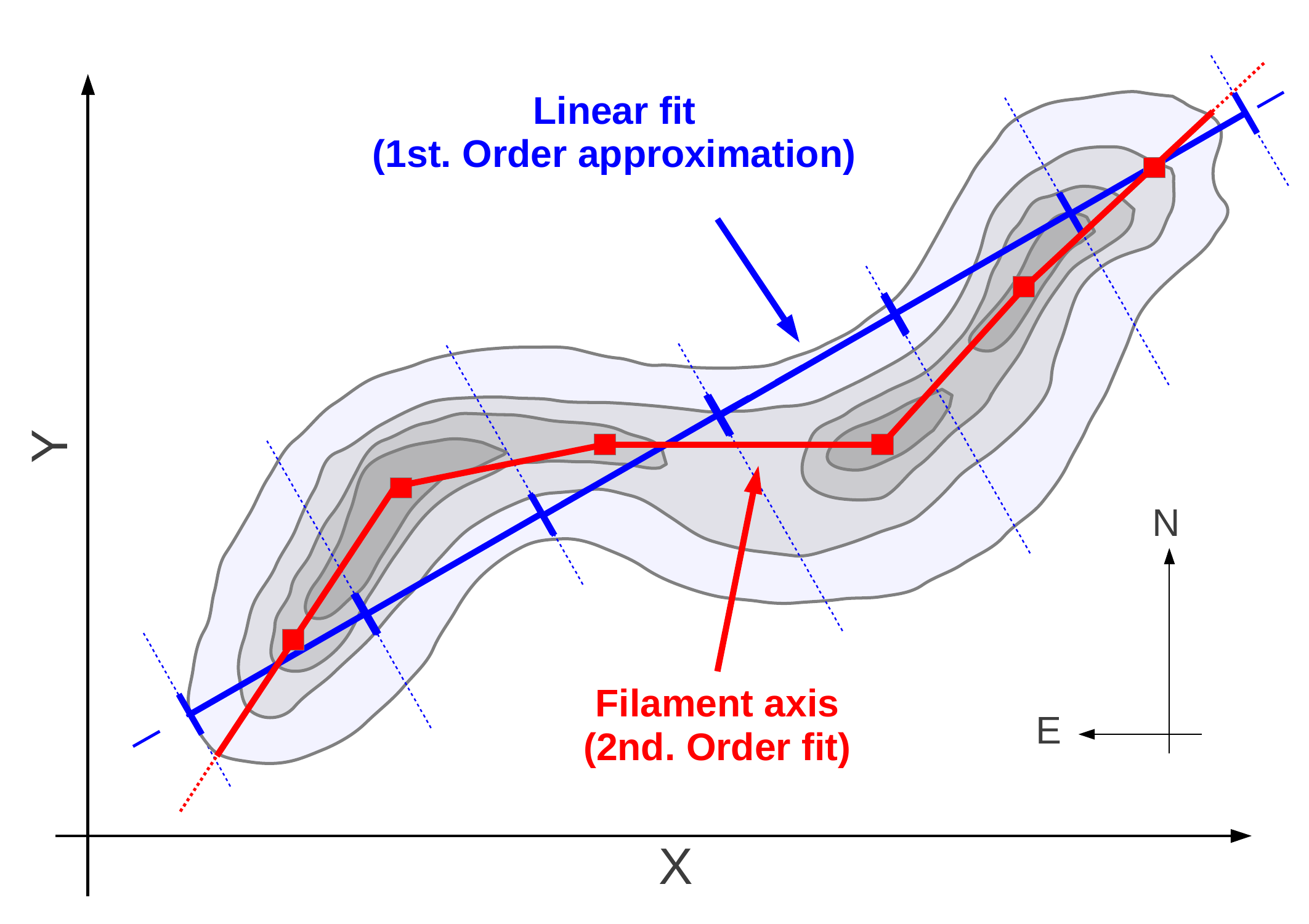}}
\caption{
Schematic view of the two-step process used to
determine the principal axis a filamentary velocity component.
The grey-scale and contours represent the C$^{18}$O(1--0) emission.
The blue straight line is the first-step determination of 
the principal axis, and the red polygonal curve is the second-step
determination of the axis.
\label{fig-axis}}
\end{center}
\end{figure}

To analyze different properties of the filamentary velocity components
identified with the {\tt FIVe} algorithm, such as their spatial location 
and velocity field, it is convenient
to determine for each of them an axis that follows the long
dimension of the filament. 
We have done this using the process illustrated in
Fig.~\ref{fig-axis}, which consists of two steps.
First, we have fitted a straight line to the 
distribution of all points identified by {\tt FIVe} as belonging to
the same velocity component, weighting each point by the 
intensity of the C$^{18}$O(1--0) emission (in cases of suspected
CO depletion, the C$^{18}$O intensity has 
been corrected as
described in Sect.~\ref{sect-fof}). As a result of this step,
we have obtained a first approximation to the 
principal axis of each velocity component 
(blue line in Fig.~\ref{fig-axis}).
Using this axis as a reference, we have then divided the set of 
points into $90''$-long segments, and for each segment, we
have determined the emission centroid following the same 
wighting scheme as in the first step.
Connecting the emission centroids with a 
polygonal curve, we have defined the
second and final approximation to the 
principal axis of each filament
(red line in Fig.~\ref{fig-axis}).

\begin{acknowledgements}
We thank Mark Heyer for assistance during the FCRAO observations
We thank Carlos De Breuck, Thomas Stanke, and Giorgio Siringo
for assistance during the APEX observations, and Axel Weiss
and Arnaud Belloche for help with the data reduction.
We also thank Guillermo Quintana-Lacaci and the IRAM staff
for help during the MAMBO observations.
An anonymous referee provided a number of useful comments and
suggestions that are greatly appreciated.
This publication is supported by the
Austrian Science Fund (FWF).
This research has made use of NASA's Astrophysics Data System 
Bibliographic Services and the SIMBAD database,
operated at CDS, Strasbourg, France.
MT acknowledges support by MINECO within the program
CONSOLIDER INGENIO 2010, under grant ``Molecular Astrophysics:
The Herschel and ALMA era - ASTROMOL'' (ref.: CSD2009-00038).
\end{acknowledgements}


\begin{thebibliography}{}

\bibitem[Alves et al.(2007)]{alv07} Alves, J., Lombardi, M., \& Lada, C.~J.\ 
2007, \aap, 462, L17 

\bibitem[Arzoumanian et al.(2011)]{arz11} Arzoumanian, D., Andr{\'e}, P., Didelon, P., et al.\ 2011, \aap, 529, L6 

\bibitem[Andr{\'e} et al.(2010)]{and10} Andr{\'e}, P., Men'shchikov, A., 
Bontemps, S., et al.\ 2010, \aap, 518, L102 

\bibitem[Barnard(1907)]{bar07} Barnard, E.~E.\ 1907, \apj, 25, 218 

\bibitem[Barnard(1927)]{bar27} Barnard, E.~E.\ 1927, 
Catalogue of 349 dark objects in the sky. (Chicago: University of Chicago
Press)

\bibitem[Belloche et al.(2011)]{bel11} Belloche, A., Schuller, F., Parise, B., 
et al.\ 2011, \aap, 527, A145 

\bibitem[Benson \& Myers(1989)]{ben89} Benson, P.~J., \& Myers, P.~C.\ 1989, \apjs, 71, 89 

\bibitem[Bergin et al.(2002)]{ber02} Bergin, E.~A., Alves,
J., Huard, T., \& Lada, C.~J.\ 2002, \apjl, 570, L101

\bibitem[Bergin et al.(2004)]{ber04} Bergin, E.~A., Hartmann, 
L.~W., Raymond, J.~C., \& Ballesteros-Paredes, J.\ 2004, \apj, 612, 921 

\bibitem[Bergin \& Tafalla(2007)]{ber07} Bergin, E.~A., \& Tafalla, M.\ 2007, \araa, 45, 339 


\bibitem[Berlind et al.(2006)]{ber06} Berlind, A.~A., 
Frieman, J., Weinberg, D.~H., et al.\ 2006, \apjs, 167, 1 

\bibitem[Bertout et al.(2007)]{bert07} Bertout, C., Siess, L., \& Cabrit, 
S.\ 2007, \aap, 473, L21 

\bibitem[Caselli et al.(2002)]{cas02} Caselli, P., Benson, 
 P.~J., Myers, P.~C., \& Tafalla, M.\ 2002, \apj, 572, 238 

\bibitem[Caselli et al.(1999)]{cas99} Caselli, P., Walmsley,
C.~M., Tafalla, M., Dore, L., \& Myers, P.~C.\ 1999, \apjl, 523, L165

\bibitem[Cernicharo et al.(1985)]{cer85} Cernicharo, J., Bachiller, R., 
\& Duvert, G.\ 1985, \aap, 149, 273 

\bibitem[Clark et al.(1977)]{cla77} Clark, F.~O., Giguere, 
P.~T., \& Crutcher, R.~M.\ 1977, \apj, 215, 511 

\bibitem[Clark et al.(2012)]{cla12} Clark, P.~C., Glover, 
S.~C.~O., Klessen, R.~S., \& Bonnell, I.~A.\ 2012, \mnras, 424, 2599 

\bibitem[Davis et al.(2010)]{dav10} Davis, C.~J., 
Chrysostomou, A., Hatchell, J., et al.\ 2010, \mnras, 405, 759 

\bibitem[di Francesco et al.(2007)]{dif07} di Francesco, J., 
Evans, N.~J., II, Caselli, P., et al.\ 2007, Protostars and Planets V, 17


\bibitem[Duvert et al.(1986)]{duv86} Duvert, G., Cernicharo, J., \& Baudry, A.\ 1986, 
\aap, 164, 349 

\bibitem[Elias(1978)]{eli78} Elias, J.~H.\ 1978, \apj, 224, 
857 

\bibitem[Elmegreen(1993)]{elm93} Elmegreen, B.~G.\ 1993, 
\apjl, 419, L29 

\bibitem[Enoch et al.(2007)]{eno07} Enoch, M.~L., Glenn, J., 
Evans, N.~J., II, et al.\ 2007, \apj, 666, 982

\bibitem[Evans(2008)]{eva08} Evans, N.~J., II 2008, Pathways 
Through an Eclectic Universe, eds. J. H.
Knapen, T. J. Mahoney, \& A. Vazdekis, ASP Conf. Ser., 390, 52


\bibitem[Frerking et al.(1982)]{fre82} Frerking, M.~A., 
Langer, W.~D., \& Wilson, R.~W.\ 1982, \apj, 262, 590 

\bibitem[Gaida et al.(1984)]{gai84} Gaida, M., Ungerechts, H., \& Winnewisser, G.\ 1984, 
\aap, 137, 17 

\bibitem[Goldsmith et al.(2008)]{gol08} Goldsmith, P.~F., 
Heyer, M., Narayanan, G., et al.\ 2008, \apj, 680, 428 

\bibitem[Goodman et al.(1998)]{goo98} Goodman, A.~A.,
Barranco, J.~A., Wilner, D.~J., \& Heyer, M.~H.\ 1998, \apj, 504, 223

\bibitem[Hacar(2012)]{hac12} Hacar, A.\ 2012, Ph.D. Thesis, Univ. Complutense, 
Madrid

\bibitem[Hacar \& Tafalla(2011)]{hac11} Hacar, A., \& Tafalla, M.\ 2011, 
\aap, 533, A34  

\bibitem[Hartmann(2002)]{har02} Hartmann, L.\ 2002, \apj,
578, 914

\bibitem[Heiles \& Katz(1976)]{hei76} Heiles, C., \& Katz, G.\ 1976, \aj, 81, 37 

\bibitem[Heitsch et al.(2008)]{hei08} Heitsch, F., Hartmann, 
L.~W., Slyz, A.~D., Devriendt, J.~E.~G., 
\& Burkert, A.\ 2008, \apj, 674, 316 

\bibitem[Heyer et al.(1987)]{hey87} Heyer, M.~H., Vrba, 
F.~J., Snell, R.~L., et al.\ 1987, \apj, 321, 855 

\bibitem[Hennebelle et al.(2008)]{hen08} Hennebelle, P., Banerjee, R., 
V{\'a}zquez-Semadeni, E., Klessen, R.~S., \& Audit, E.\ 2008, \aap, 486, L43 

\bibitem[Hirota et al.(2002)]{hir02} Hirota, T., Ito, T., 
\& Yamamoto, S.\ 2002, \apj, 565, 359 

\bibitem[Hirota et al.(2004)]{hir04} Hirota, T., Maezawa, H., 
\& Yamamoto, S.\ 2004, \apj, 617, 399 

\bibitem[Huchra \& Geller(1982)]{huc82} Huchra, J.~P., \& Geller, M.~J.\ 1982, 
\apj, 257, 423 

\bibitem[Johnstone \& Bally(1999)]{joh99} Johnstone, D., \& Bally, J.\ 1999, 
\apjl, 510, L49 


\bibitem[Kenyon et al.(2008)]{ken08} Kenyon, S.~J.,
G{\'o}mez, M., \& Whitney, B.~A.\ 2008, Handbook of Star Forming
Regions, Volume I, ed. B. Reipurth, 405

\bibitem[Klessen et al.(2005)]{kle05} Klessen, R.~S.,
Ballesteros-Paredes, J., V{\'a}zquez-Semadeni, E.,
\& Dur{\'a}n-Rojas, C.\ 2005, \apj, 620, 786

\bibitem[Kraus \& Hillenbrand(2009)]{kra09} Kraus, A.~L., \& 
Hillenbrand, L.~A.\ 2009, \apj, 704, 531


\bibitem[Larson(1981)]{lar81} Larson, R.~B.\ 1981, \mnras,
194, 809

\bibitem[Larson(1995)]{lar95} Larson, R.~B.\ 1995, \mnras, 
272, 213 



\bibitem[Leung(1978)]{leu78} Leung, C.~M.\ 1978, \apj, 225, 
427 

\bibitem[Li \& Goldsmith(2012)]{li12} Li, D., \& Goldsmith, P.~F.\ 2012, \apj, 756, 12 

\bibitem[Luhman et al.(2009)]{luh09} Luhman, K.~L., Mamajek, 
E.~E., Allen, P.~R., \& Cruz, K.~L.\ 2009, \apj, 703, 399 

\bibitem[Lynds(1962)]{lyn62} Lynds, B.~T.\ 1962, \apjs, 7, 1 

\bibitem[Men'shchikov et al.(2012)]{men12} Men'shchikov, A., Andr{\'e}, P., 
Didelon, P., et al.\ 2012, \aap, 542, A81

\bibitem[Mizuno et al.(1995)]{miz95} Mizuno, A., Onishi, T., 
Yonekura, Y., et al.\ 1995, \apjl, 445, L161 


\bibitem[Molinari et al.(2010)]{mol10} Molinari, S., Swinyard, B., Bally, J., et al.\ 2010, 
\aap, 518, L100 

\bibitem[Motte et al.(1998)]{mot98} Motte, F., Andre, P., \& Neri, R.\ 
1998, \aap, 336, 150 

\bibitem[Mouschovias \& Ciolek(1999)]{mou99} Mouschovias, T.~C.,
\& Ciolek, G.~E.\ 1999, NATO ASIC Proc.~540: The Origin of Stars and
Planetary Systems, 305

\bibitem[Myers(2009)]{mye09} Myers, P.~C.\ 2009, \apj, 700,
1609

\bibitem[Narayanan et al.(2008)]{nar08} Narayanan, G., Heyer, 
M.~H., Brunt, C., et al.\ 2008, \apjs, 177, 341 

\bibitem[Onishi et al.(1996)]{oni96} Onishi, T., Mizuno, A., 
Kawamura, A., Ogawa, H., \& Fukui, Y.\ 1996, \apj, 465, 815 

\bibitem[Onishi et al.(2002)]{oni02} Onishi, T., Mizuno, A., 
Kawamura, A., Tachihara, K., \& Fukui, Y.\ 2002, \apj, 575, 950 

\bibitem[Ostriker(1964)]{ost64} Ostriker, J.\ 1964, \apj,
140, 1056

\bibitem[Padoan et al.(2001)]{pad01} Padoan, P., Juvela, M.,
Goodman, A.~A., \& Nordlund, {\AA}.\ 2001, \apj, 553, 227

\bibitem[Palau et al.(2012)]{pal12} Palau, A., de 
Gregorio-Monsalvo, I., Morata, {\`O}., et al.\ 2012, \mnras, 424, 2778

\bibitem[Palmeirim et al.(2013)]{pal13} Palmeirim, P., Andr{\'e}, P., Kirk, J., 
et al.\ 2013, \aap, 550, A38 

\bibitem[Rebull et al.(2010)]{reb10} Rebull, L.~M., Padgett, 
D.~L., McCabe, C.-E., et al.\ 2010, \apjs, 186, 259 

\bibitem[Schmalzl et al.(2010)]{sch10} Schmalzl, M., 
Kainulainen, J., Quanz, S.~P., et al.\ 2010, \apj, 725, 1327 

\bibitem[Schneider \& Elmegreen(1979)]{sch79} Schneider, S., \&
Elmegreen, B.~G.\ 1979, \apjs, 41, 87

\bibitem[Simon(1997)]{sim97} Simon, M.\ 1997, \apjl, 482, L81 

\bibitem[Siringo et al.(2009)]{sir09} Siringo, G., Kreysa, E., Kov{\'a}cs, A., et
al.\ 2009, \aap, 497, 945

\bibitem[Shu et al.(1987)]{shu87} Shu, F.~H., Adams, F.~C.,
\& Lizano, S.\ 1987, \araa, 25, 23

\bibitem[Smith et al.(2012)]{smi12} Smith, R.~J., Shetty, R., 
Stutz, A.~M., \& Klessen, R.~S.\ 2012, \apj, 750, 64 

\bibitem[Stod{\'o}lkiewicz(1963)]{sto63} Stod{\'o}lkiewicz,
J.~S.\ 1963, Acta Astronomica, 13, 30


\bibitem[Tafalla et al.(2002)]{taf02} Tafalla, M., Myers,
P.~C., Caselli, P., Walmsley, C.~M., \& Comito, C.\ 2002, \apj, 569,
815

\bibitem[Tafalla et al.(2004)]{taf04a} Tafalla, M., Myers, P.~C., Caselli, P., \& Walmsley, C.~M.\ 
2004, \aap, 416, 191

\bibitem[Tafalla \& Santiago(2004)]{taf04b} Tafalla, M., \& Santiago, J.\ 2004, \aap, 414, L53 

\bibitem[Tafalla et al.(2006)]{taf06} Tafalla, M., Santiago-Garc{\'{\i}}a, J., 
Myers, P.~C., et al.\ 2006, \aap, 455, 577 


\bibitem[Tatematsu et al.(2004)]{tat04} Tatematsu, K., Umemoto, T., Kandori, R., 
\& Sekimoto, Y.\ 2004, \apj, 606, 333 

\bibitem[V{\'a}zquez-Semadeni et al.(2005)]{vaz05}
V{\'a}zquez-Semadeni, E., Kim, J., Shadmehri, M.,
\& Ballesteros-Paredes, J.\ 2005, \apj, 618, 344

\bibitem[V{\'a}zquez-Semadeni et al.(2006)]{vaz06} 
V{\'a}zquez-Semadeni, E., Ryu, D., Passot, T., Gonz{\'a}lez, 
R.~F., \& Gazol, A.\ 2006, \apj, 643, 245 

\bibitem[Ward-Thompson et al.(2007)]{war07} Ward-Thompson, 
D., Andr{\'e}, P., Crutcher, R., et al.\ 2007, Protostars and Planets V, 33 

\end{thebibliography}
\end{document}